\documentclass[11pt]{article}
\usepackage{setspace}
\usepackage{indentfirst}
\usepackage{exscale,relsize}
\usepackage{prettyref}
\usepackage{amssymb,amsthm}
\usepackage[tbtags]{amsmath}
\usepackage{array,supertabular}
\usepackage{bm}
\usepackage[]{epsfig}
\usepackage{rotating}
\usepackage{graphicx}
\usepackage{mathtools}
\usepackage{longtable}
\usepackage[round]{natbib}
\usepackage{graphicx}
\usepackage{subcaption}
\usepackage{color,colortbl}
\usepackage{pgfplots}
\usepackage{lscape}

\usepackage{amsmath}
\usepackage{pgfplots}
\usepackage{tikz}
\usepackage{mathtools,amsmath}
\pgfplotsset{compat=1.16}
\usepackage{caption}

\usepackage{tikz}
\usepackage{float}
\usepackage{mathptmx}
\usepackage[11pt]{moresize}
\usepackage[affil-it]{authblk}

\usepackage{booktabs}
\usepackage{pdflscape}

\usepackage{hyperref}                 
\hypersetup{
    pdfborder={0 0 0},                                    
    colorlinks,                                                    
    linkcolor=blue,
    citecolor=black,
    urlcolor=blue
    }

\definecolor{LightCyan}{rgb}{0.88,1,1}
\definecolor{Yellow}{rgb}{1,1,0}
\definecolor{Blue}{rgb}{0,0,0.8}
\definecolor{Red}{rgb}{1,0,0}

\newrefformat{sec}{Section~\ref{#1}}
\newrefformat{tab}{Table~\ref{#1}}

\def\fillandplacepagenumber{%
 \par\pagestyle{empty}%
 \vbox to 0pt{\vss}\vfill
 \vbox to 0pt{\baselineskip0pt
   \hbox to\linewidth{\hss}%
   \baselineskip\footskip
   \hbox to\linewidth{%
     \hfil\thepage\hfil}\vss}}

\pagebreak

\title{ \vspace*{-2.5cm} \hspace*{-0.5cm}Financing Costs, Per-Shipment Costs and Shipping Frequency: \\Firm-Level Evidence from Bangladesh \footnote{
I am grateful to Dr. Georg Schaur for all his feedback on this study. 
}}

\author{Md Deluair Hossen\thanks{University of Tennessee, Knoxville.
\href{mailto:TK@TK.edu}{mhossen@utk.edu}} }

\date{ \vspace*{0.5cm} March, 2023\\
} 

\pgfplotsset{compat=1.15}

\begin{document}

\maketitle

\begin{abstract}
In international trade, firms face lengthy ordering-producing-delivery times and make shipping frequency decisions based on the per-shipment costs and financing costs. In this paper, I develop a model of importer-exporter procurement where the importer procures international inputs from exporting firms in developing countries. The exporters are credit constrained for working capital, incur the per-shipment fixed costs, and get paid after goods are delivered to the importer. The model shows that the shipping frequency increases for high financing costs in origin and destination. Furthermore, longer delivery times increase shipping frequency as well as procurement costs. The model also shows that the higher per-shipment fixed costs reduce the shipping frequency, in line with previous literature. Reduced transaction costs lower the exporter's demand for financial services through shipping frequency adjustment, mitigating the financial frictions of the firm. Then, I empirically investigate whether the conclusions regarding the effect of per-shipment fixed costs on shipping frequency from the theoretical model and in the existing literature extend to developing countries. My estimation method addresses several biases. First, I deal with aggregation bias with the firm, product, and country-level analysis. Second, I consider the Poisson Pseudo Maximum Likelihood (PPML) estimation method to deal with heteroscedasticity bias from the OLS estimation of log-linear models. Third, I fix the distance non-linearity of Bangladeshi exports. Finally, I consider the effect of financing cost on shipping frequency to address omitted variable bias. Using transaction-level export data from Bangladesh, I find that 10\% higher per-shipment costs reduce the shipping frequency by 3.45\%. This shipping frequency response is considerably higher than that of developed countries. The findings are robust to different specifications and subsamples. 

\end{abstract}

\section{Introduction}

Ordering, producing, and shipping products in international trade take considerable time \citep{anderson2004trade,hummels2013time}. This creates some challenges. Firms engaged in international trade must determine the optimal order and shipping frequency to minimize procurement costs. Long-time lags are costly for suppliers in developing countries because standard trade financing contracts require them to borrow working capital for longer. Furthermore, ICC Global Survey on Trade Finance, 2018, reports a \$1.5 trillion trade finance gap between demand and supply. This shortage of capital creates credit constraints for firms located in a country with weak financial conditions \citep{icc2018}. These problems are discussed in two disconnected works of literature. The shipping frequency literature focuses on the trade-off between the per-shipment fixed costs, and inventory storage costs \citep{alessandria2010great,alessandria2010inventories, alessandria2011us,kropf2014fixed,hornok2015administrative}. The financial constraints literature focuses on interest costs, types of financial contracts, and firm heterogeneity \citep{amiti2011exports, manova2012credit,antras2015poultry}. 

In the theoretical part of the paper, I develop a model that combines these two kinds of literature. In the model, the importer procures a product from a financially constrained exporter in a developing country. The exporter gets paid after delivery and requires working capital financing for production. A long delivery time creates a trade-off between the variable financing costs and the per-shipment fixed costs for determining optimal shipping size. An increase in delivery time raises the marginal procurement cost of order size relative to the marginal cost of placing an additional shipment. In my model, this has implications for the effects of financing costs and delivery times on order frequency and size. Moreover, the effects of per-shipment costs on the financing demand are stronger in financially constrained countries.

This paper contributes to the buyer-seller procurement contract literature \citep{taylor1997competition,bajari1999procurement,bajari2001incentives,liker2004building,antras2004global,nunn2007relationship,heise2015trade}. In particular, this paper is closely related with \cite{taylor1997competition}, which solves the shipping frequency optimization problem when the buyer is sourcing their procurement from the sellers with a per-shipment fixed inspection cost. The authors optimize order size for two distinct types of buyer-suppliers relations, what they term Japanese and American procurement systems. On the one hand, I am simplifying their model by ignoring inspections. On the other hand, I am focusing on the problem of delivery times and asymmetric interest rates in an international trade setting. My extension makes several contributions to the literature.

My model shows how asymmetric financing costs affect shipping frequencies in the importer and exporter countries. Higher exporters' borrowing rates increase financing costs as well as shipping frequency. Greater discounting in the importer country leads to more frequent shipments as delaying immediate costs over small shipments is cheaper. This contributes to the financial constraint literature \citep{amiti2011exports,chor2012off,manova2012credit,wagner2014credit, feenstra2014exports}. In this context, \cite{manova2012credit} focuses on the financial development of countries and the financial vulnerability of different sectors. In their model, the exogenous financial constraints on the fixed costs determine trade volume, whereas, in my model, exogenous financing costs on the variable costs determine shipping size. Illustrating how shipping size adapts to financial constraints, as I do in this study, is an important addition to this literature.

With the gradual decline of tariff barriers, recent studies focus on the per-shipment fixed costs, which include order processing, monitoring, customs clearance, and sanitary checking costs \citep{alessandria2010inventories, hornok2015per,bekes2017shipment, zhao2019export}. The trade-off between the variable inventory costs and the per-shipment fixed costs determines optimal shipping frequency and explains lumpy, infrequent trade. However, these studies do not consider the financing costs for determining shipping frequency. For example, \cite{hornok2015administrative} shows that the demand-side time preference of consumers and the per-shipment fixed costs determine the frequency and size of the shipment. \cite{kropf2014fixed} considers the firm's trade-off between the per-shipment fixed costs and the variable storage costs at the destination for shipping frequency optimization. By considering the financing cost's effect on shipping frequency, I show how financing the per-shipment fixed costs later in the supply chain can significantly affect the exporter's financing demand.

My model shows that the effect of longer delivery times on trade costs depends on financing and per-shipment costs. The intuition is that longer delivery times increase procurement costs, which increase even more for higher per-shipment costs. A long literature examines the effects of delivery times on trade costs \citep{djankov2010trading,berman2012time,hummels2013time,martincus2015customs,bourgeon2016financing,li2019time}. This literature identifies time as a trade cost and investigates the sources of these costs. My model contributes to identifying the nexus between delivery time, financing costs, and per-shipment costs as a potential source of the time cost. Furthermore, the time costs literature speculates on how financing costs may be a driving factor for times costs. My model shows that shipping frequencies are important to understand this mechanism, and that has been ignored.

Recent trade finance literature considers payment contract tools and financing methods to understand trade patterns \citep{schmidt2013towards,glady2011bank,ahn2014understanding,antras2015poultry,niepmann2017no}. This literature does not consider the effect of the shipping frequency on the trade flow. The shipping frequency decisions can mitigate the demand for trade finance and should be endogenized to get the full effect of trade finance on trade patterns. Regarding trade finance, my model is similar to the model developed by \cite{schmidt2013towards}, which considers the financing cost of the origin and destination, contract enforceability, and delivery time to predict the flow of trade. Using the variables above and abstracting away from trade flow, my model endogenizes the shipping frequency decision. Furthermore, the lag of international trade and trade finance mode are intertwined with the shipping frequency decisions. 

My model also shows that higher per-shipment costs lead to greater financing demand. The intuition is that higher per-shipment costs increase total fixed costs, which reduce shipping frequency but increase shipping size. The working capital for a large shipment increases the financing demand. Conversely, low per-shipment costs smooth out demands for financial services through multiple shipments. This proposition is policy-relevant. Recent trade policy focuses on trade facilitation which has large effects on per-shipment costs \citep{hoekman2010assessing,moise2013trade,carballo2016transit,carballo2018transportation}. 

In the empirical part of the paper, I consider the fact that, in international trade, shipping frequency is important to understand procurement costs \citep{taylor1997competition}, inventory adjustment \citep{alessandria2010inventories,alessandria2011us}, pricing to market and exchange rate pass-through \citep{aizenman2004endogenous}, the transmission of shocks across markets \citep{bekes2017shipment}, and effects of trade facilitation policies \citep{carballo2016transit}. Furthermore, I test the effect of per-shipment fixed costs on shipping frequency, a theoretical outcome from the model. Recent evidence examines the determinants of shipping frequency. In particular, for developed countries using product and country-level data, \cite{hornok2015administrative,hornok2015per} provide evidence that per-shipment fixed costs are an important determinant to understanding shipping frequencies. To date, evidence for developing countries using firm-level data does not exist.

There are several reasons why the existing evidence might not immediately extend to developing countries, such as Bangladesh, which is my focus. First, at the product and country level, estimates might be subject to aggregation bias \citep{lewbel1992aggregation,van2000cross,albuquerque2003practical}. Second, the OLS estimation of log-linear models in the existing literature might be biased due to heteroscedasticity \citep{silva2006log}. Third, existing specifications do not consider financing costs and might therefore be subject to omitted variable bias \citep{taylor1997competition}. Fourth, Bangladesh's unique position in the supply chain and focus on apparel products with main target markets in Europe and the US suggest that existing specifications are inappropriate for understanding shipping frequency determinants. For example, given the geographic location of Bangladesh, it is not clear that shipping frequencies are linearly decreasing in distance, as a common assumption of existing specifications. In this paper, my research question is if the conclusions regarding per-shipment fixed costs in the existing literature extend to developing countries.

To answer this research question, I employ transaction-level export data from Bangladesh from 2006 to 2013. Consistent with the existing literature, I measure shipping frequency as the number of export shipments a firm in Bangladesh engages in with foreign buyers for a given product within a year. I apply robust estimators appropriate for count models concerning the heteroscedasticity bias in log-linear models. I account for non-linear distance effects in my empirical model and provide evidence for aggregation bias considering product, country, and mode of transportation levels of aggregation. To examine omitted variable bias, I extend the model for various measures of financing costs and take advantage of my firm-level data to specific unobserved heterogeneity with fixed effects.

My main findings are that OLS estimation of the log-linear model of \cite{hornok2015administrative, hornok2015per} underestimates the per-shipment cost effects on the shipping frequency at both the country and product levels. The Poisson Pseudo Maximum Likelihood (PPML) estimation method can overcome this underestimation. I find that the per-shipment cost estimates are overestimated at the higher levels of aggregation. I also find evidence of omitted variable bias as adding financing cost increases the elasticity of the per-shipment costs. Finally, I find that for Bangladeshi trade, fixing distance non-linearity is crucial to understand the effect of per-shipment costs on the shipping frequency. By fixing all of these biases, I find that for a 10\% increase in the per-shipment costs, the shipping frequency decreases by $3.24$\% for Bangladesh. By comparing the findings of \cite{hornok2015per}, this shipping frequency reduction is 1.28 times larger than the US and 3.22 times larger than Spain.

My empirical results provide four contributions to the existing literature.

First, by solving various identification problems in the existing literature, my evidence shows that per-shipment fixed costs are more important in developing countries relative to developed countries. At the exporter level, the per-shipment elasticity is -0.227 by accounting for a product-mode-year effect. This elasticity is higher than the findings of \cite{hornok2015per} for the product-level Spanish exports.

Second, my evidence shows that applications in the existing literature are subject to biases. One such relevant bias is omitted variable bias. When I introduce financing costs and narrow the fixed effects to account for unobserved heterogeneity, the per-shipment cost elasticity increased by 1.48 times. Moreover, the effect of financing costs on shipping frequency remains robust to additional measures of financing costs. Furthermore, addressing omitted variable problems also increases the effects of country size (GDP), income (GDP per capita), and distance on the shipping frequency.

Third, financial constraints are an important determinant of trade flows and a firm's selection into export markets \citep{muuls2008exporters,bellone2010financial, amiti2011exports,chor2012off,manova2012credit,buono2013bank,besedevs2014export,paravisini2014dissecting,fauceglia2015credit,bergin2017firm}. Existing procurement models show that if there is heterogeneity in financing costs across exporter and importer locations, then financing costs are also an important determinant of shipping frequencies \citep{taylor1997competition,schmidt2013towards}. Understanding if financing costs are a determinant of shipping frequencies is important because if they are, then shipping frequencies affect the demand for financial services, and policies that reduce per-shipment fixed costs have the potential to relax financial constraints. Consistent with the predictions from these models, I provide evidence that greater exporter interest rates increase shipping frequency more for higher importer discount rates. However, higher importer discount rates decrease shipping frequency, contradicting the theoretical predictions. Overall, fixed cost-reducing policies can decrease the severity of financial constraints more in high-interest rate countries than low-interest rate countries by increasing shipping frequency.

Finally, the WTO agreement on trade facilitation strives to reduce transaction barriers in international trade. Reducing trade costs in a developing country is a key goal of the Aid for Trade \citep{wto2013} and trade facilitation programs \citep{wto2014}. Evidence shows that these policies reduce fixed-per-shipment costs \citep{moise2013trade, goldberg2016effects,carballo2016border, carballo2016transit, carballo2018transportation}. For example, Bangladesh, a developing country, continues to reduce trade costs as a beneficiary country of the Trade Facilitation Support Program by the World Bank. This study shows that these policies reduce trade costs and relax the demand for financial services in financially constrained countries. This provides a mechanism for evaluating the influence of trade policies. Based on my estimates, these policies might be more effective for developing countries like Bangladesh than developed countries. Furthermore, these policies are especially important compared to developed countries in determining exchange rate pass-through, inventory adjustment, and the transmission of shocks across markets due to increased shipping frequencies.

Having discussed the background and related literature, the remainder of this paper is organized as follows. Section 2 builds up the model and explains the theoretical outcomes with propositions. Section 3 describes the data and estimation strategies. Section 4 presents the results, and section 5 checks robustness with several subsamples and alternative specifications. Section 6 concludes.

\section{The Model Setup}
I develop a simple model of the exporter-importer procurement system with financial constraints and delivery times. An importer procures quantity $q$ and minimizes cost by choosing an order size $x$, which implies that the shipping frequency is $n=\frac{q}{x}$. I treat the choice of order size and shipping frequency as exogenous while focusing on the exporter's problem. The importer pays after receiving the order. This assumption is consistent with standard trade financing contracts literature \citep{schmidt2013towards,niepmann2017international}. These payments after delivery create a financing problem for the exporter.\footnote{The exporter financing due to payment after delivery is more common in developing countries. For example, export financing is used for 44.36 \% of Bangladeshi exports by volume \citep{habib2017trade}.}

\subsection{The Exporter's Problem}
The exporter receives an order of size $x.$ The exporter's constant marginal cost of production is $c.$ Therefore, the production cost is $cx.$ Located in a developing country, the exporter, is financially constrained and borrows working capital from domestic financial intermediaries to cover production and other costs \citep{chan2014financial}. Let $\Delta$ be the time between the importer's order and the goods delivered, defined here as delivery time. This includes time for production, domestic transportation to the port, customs clearance, and port-to-port transportation. Assume $\Delta$ is independent of order size $x$ as most of the delivery time's components are fixed except production time. For example, a garment shipment from a firm in Bangladesh to JC Penny in the USA may take 116 days from getting the order to the product delivered, whereas production time is only 23 days \citep{haque2009lead}. Let $r$ be the interest at which the exporter borrows the entire working capital for $\Delta.$ With the continuous compounded interest rate, a \$1 borrowing requires $\big (\mathrm{e}^{\Delta r}\big)$ payment at the end of $\Delta$ period. The total production cost, including financing cost, is then $\big (c x \mathrm{e}^{\Delta r}\big).$ 

After production, the firm ships the goods to the importer. Here, I am modeling the per-shipment fixed costs as if they appear later in the supply chain, which does not require upfront borrowing \citep{barry2016basic,schutte2019post}. I am considering this because it is a more interesting case to understand the trade-off between the financing cost and the per-shipment costs for determining the shipping frequency.\footnote{Regarding the per-shipment fixed costs for shipping $f$, there are two possible assumptions. The per-shipment costs are borrowed immediately or paid later in the supply chain and do not require borrowing. I show the first case in Appendix-\hyperref[a1]{A}.} With the per-shipment cost, the total cost of producing and delivering an individual shipment is $(c x \mathrm{e}^{\Delta r}+ f)$.

The exporter is willing to produce and deliver if the payment from the importer $w(x)$ at least covers the total cost to produce and ship the order. Therefore, the exporter's participation constraint is
 \begin{equation*}
     w(x)\geq cx\mathrm{e}^{\Delta r} + f
 \end{equation*}
 
\subsection{The Importer's Problem}
I derive an importer's cost for a single order in this section. I then aggregate all the orders to get the total procurement cost and optimize the order size. Finally, I derive propositions. 

With all the bargaining power, the importer covers costs only (i.e., pays no rent to the exporter). Moreover, with the constant marginal cost and competitive environment, the exporter's participation is ensured even if the payment only covers the costs. The exporter participation gives 
\begin{equation}\label{eq:1}
    w(x)= c x \mathrm{e}^{ \Delta r}+ f 
\end{equation}

Equation \ref{eq:1} shows that if exporters are financially constrained and delivery times are long, importers need to make a greater payment for procurement to cover higher financing costs.

As the importer pays for the order after receiving the shipment, the importer's present value of cost for this single shipment is
\begin{equation*}
    \mathrm{e}^{- \Delta r_1}w(x)
\end{equation*} 

Here, $r_1$ is the importer's domestic annual interest rate, which is assumed to be independent of $\Delta$. Furthermore, the importer discounts the procurement costs of a single order by $(\mathrm{e}^{- \Delta r_1})$.

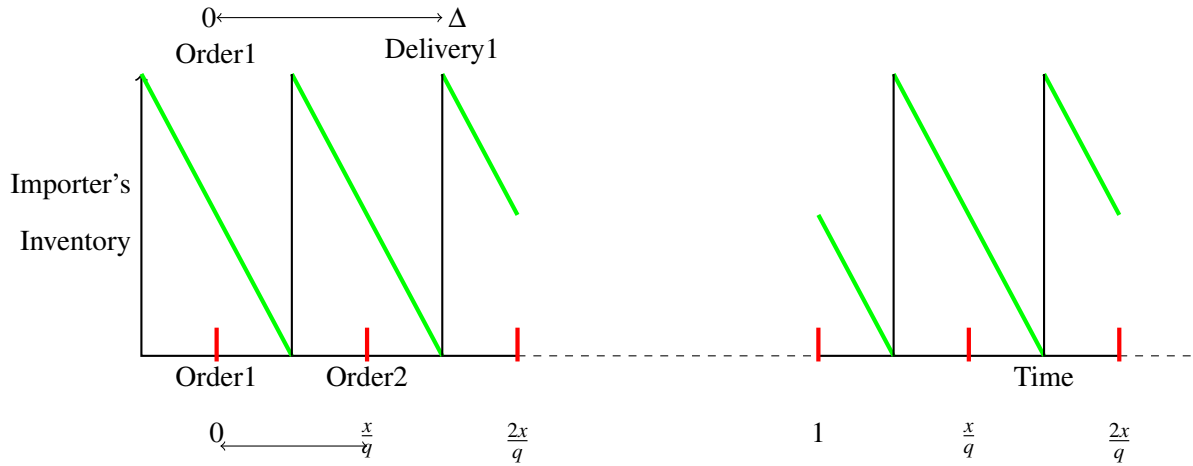
\begin{figure}[H]
    \centering
    \renewcommand\thefigure{1}
    \begin{tikzpicture}[xscale=10,yscale=15]
\draw [thick, <-] (0,0.25) -- (0,0) -- (0.5,0);
\draw [dashed, .] (0.5,0) -- (0.9,0);
\draw [thick, -] (0.9,0) -- (1.3,0);
\draw [dashed, .] (1.3,0) -- (1.4,0);

\draw[green, ultra thick](0,0.25) -- (0.2,0)  ;
\draw[green, ultra thick](0.2,0.25) -- (0.4,0)  ;
\draw[green, ultra thick](0.4,0.25) -- (0.5,0.125)  ;
\draw[green, ultra thick](0.9,0.125) -- (1,0)  ;
\draw[green, ultra thick](1,0.25) -- (1.2,0)  ;
\draw[green, ultra thick](1.2,0.25) -- (1.3,0.125)  ;

\draw[red, ultra thick](0.1,0.025) -- (0.1,-0.005)  ;
\draw[red, ultra thick](0.3,0.025) -- (0.3,-0.005)  ;
\draw[red, ultra thick](0.5,0.025) -- (0.5,-0.005)  ;
\draw[red, ultra thick](0.9,0.025) -- (0.9,-0.005)  ;
\draw[red, ultra thick](1.1,0.025) -- (1.1,-0.005)  ;
\draw[red, ultra thick](1.3,0.025) -- (1.3,-0.005)  ;

\draw [thick, -] (0.2,0.25) -- (0.2,0);
\draw [thick, -] (0.4,0.25) -- (0.4,0);
\draw [thick, -] (1,0.25) -- (1,0);
\draw [thick, -] (1.2,0.25) -- (1.2,0);
\draw [ <->] (0.1,0.3) -- (0.4,0.3);
\draw [ <->] (0.105,-0.08) -- (0.3,-0.08);

\node [below] at (0.1,0) {Order1};
\node [below] at (0.3,0) {Order2};
\node [below] at (0.1,-0.05) {0};
\node [below] at (0.3,-0.05) {$\frac{x}{q}$};
\node [below] at (0.5,-0.05) {$\frac{2x}{q}$};
\node [below] at (1.1,-0.05) {$\frac{x}{q}$};
\node [below] at (1.3,-0.05) {$\frac{2x}{q}$};
\node [below] at (0.9,-0.05) {1};
\node [below] at (1.2,0) {Time};
\node [above] at (0.1,0.25) {Order1};
\node [above] at (0.4,0.25) {Delivery1};
\node  at (0.09,0.3) {0};
\node  at (0.42,0.3) {$\Delta$};
\node [left] at (0,0.15) {Importer's };
\node [left] at (0,0.10) {Inventory};

\end{tikzpicture}

    \caption{Ordering Process in International Trade }
    \label{dt}
\end{figure}

The importer buys in batches of the same size $x,$ similarly spaced throughout the entire period. Figure \ref{dt} shows a case where the delivery time is larger than the time between two consecutive orders. It takes $\Delta$ time from the first order to the first delivery. Although goods arrive after $\Delta$ time after the order is placed, the buyer still gets products $\frac{x}{q}$ time apart in his warehouse. The importer discounts each order cost for the time duration of $\frac{x}{q}.$ Infinite numbers of such discounted single-order costs are added to get the importer's aggregated present procurement costs of the indefinite supply stream. Thus, the total procurement cost of $C$ is

\begin{equation*}
    C = w(x)\mathrm{e}^{-\Delta r_1}(1+\mathrm{e}^{-\frac{r_1x}{q}}+\mathrm{e}^{-\frac{2r_1x}{q}}+.... ) = \frac{w(x)\mathrm{e}^{-\Delta r_1}}{1-\mathrm{e}^{-\frac{r_1x}{q}}}
\end{equation*}

The importer chooses the optimal shipment size to minimize costs. 
\begin{equation} \label{eq:2}
    \min_{x} \frac{w(x)\mathrm{e}^{-\Delta r_1}}{1-\mathrm{e}^{-\frac{r_1x}{q}}} 
\end{equation}

Substituting $w(x)$ from equation \ref{eq:1} into equation \ref{eq:2} yields following optimization problem with asymmetric interest rates\footnote{ Interest rates segmentation \citep{mishkin1984real} and the capital market-deepening \citep{rajan1998financial} ensure asymmetric interest rates for any particular exporter-importer country pair $(r\neq r_1).$}  
\begin{equation*}
   \min_{x} \dfrac{(cx\mathrm{e}^{ \Delta r}+f)\mathrm{e}^{-\Delta r_1}}{1-\mathrm{e}^{-\frac{r_1x}{q}}}
\end{equation*}

Taking derivatives of the above equation concerning $x$ and setting equal to zero yield 
\begin{equation*}
    \dfrac{c\mathrm{e}^{\Delta r}(\mathrm{e}^{-\Delta r_1})}{1-\mathrm{e}^{-\frac{r_1x}{q}}}-\dfrac{r_1(\mathrm{e}^{-\Delta r_1})(c\mathrm{e}^{\Delta r}x+f)\mathrm{e}^{-\frac{r_1x}{q}}}{q(1-\mathrm{e}^{-\frac{r_1x}{q}})^2} =0
\end{equation*}

Rearranging,
\begin{equation*}
    \bigg[\dfrac{c\mathrm{e}^{\Delta r}(\mathrm{e}^{-\Delta r_1})}{1-\mathrm{e}^{-\frac{r_1x}{q}}}-\dfrac{r_1(\mathrm{e}^{-\Delta r_1})(c\mathrm{e}^{\Delta r}x)\mathrm{e}^{-\frac{r_1x}{q}}}{q(1-\mathrm{e}^{-\frac{r_1x}{q}})^2}\bigg]-\dfrac{r_1(\mathrm{e}^{-\Delta r_1})f\mathrm{e}^{-\frac{r_1x}{q}}}{q(1-\mathrm{e}^{-\frac{r_1x}{q}})^2} =0
\end{equation*}

An increase of $x$ makes the marginal transaction cost in the second term of the above equation smaller but makes the immediate marginal purchasing costs in the square-bracketed term larger. Thus, for choosing $x$, the firm faces a trade-off between the more per-shipment fixed cost of an additional order and more immediate procurement costs of a larger shipment due to one less order.

After manipulation, the optimized order size is the root of the following equation:
\begin{equation} \label{eq:3}
    qc \mathrm{e}^{\Delta r} (\mathrm{e}^{\frac{r_1 x^*}{q}}-1) - c \mathrm{e}^{\Delta r} r_1 x^* = f r_1
\end{equation}

The equation \ref{eq:3} shows that the optimized order size is a function of financing costs, delivery times, marginal cost, and quantity. Thus, $x^* = x(r, r_1, \Delta, f, c, q).$\footnote{Considering the dependency of order size on interest rates, procurement amount, and per-shipment costs, this closely follows Baumol's model of the transactions demand for money \citep{baumol1952transactions}.} 

\textbf{Proposition 1.} 
\textit{For $x>0$, there exists a unique shipping size, $x^*$ that minimized the firm's procurement costs.}

\textbf{Proof:} See Appendix-\hyperref[a1]{A}

For positive shipping size, solving equation \ref{eq:3} yields a unique shipping size. This shipping size depends on additional variables than the optimal shipping size derived in \cite{taylor1997competition}. The origin and destination financing costs and delivery times are now crucial for the procurement's shipping size decision. The effect of delivery time on the shipping size in my model results from taking the buyer-seller domestic setup of \cite{taylor1997competition} to international trade. 

\subsubsection{Shipping Frequency}
The optimized order size yields the shipping frequency. As the shipping frequency is an important decision factor for trade, I investigate how other relevant model variables can affect the shipping frequency in the following proposition.

\textbf{Proposition 2.} 
\textit{ All else unchanged, shipping frequency} \textit{(i) increases if the interest rate in the exporter's country increases: $\frac{\partial n}{\partial r} \geq 0$,} \textit{(ii) increases if the interest rate in the importer's country increases: $\frac{\partial n}{\partial r_1} \geq 0$,} \textit{(iii) increases if the delivery time is longer: $\frac{\partial n}{\ \partial \Delta } \geq 0$}, and \textit{(iv) decreases if the per-shipment fixed costs increase: $\frac{\partial n}{\partial f } \leq 0$.}

\textbf{Proof:} See Appendix-\hyperref[a1]{A}

The intuition for the first result is as follows. First, if exporters' financing costs increase, they require a higher payment, and importers respond by increasing shipping frequency to push added costs into the future. Thus, facing the higher-interest rate in the exporter country, the importer increases the shipping frequency. Next, if financing costs increase in the importer country, importers choose small-size orders to benefit from higher discount rates. As the fixed costs are spanned throughout the entire period over smaller shipment sizes, greater discounting reduces the total costs. Finally, the financing need arises for the exporter from the delayed payment after delivery to the importer. The longer delivery times increase the exporter's financing need for an extended period, increasing financing costs. To avoid these high financing costs, the firm chooses more frequent shipments. Considering a fixed amount of procurement, anything that makes the exporter's production costly forces the importers to choose a small lot size, reducing some of the financing costs through a small time window between shipments.

The simple intuition for the last result is that high transaction costs make frequent shipments costly to the importer. The importer chooses less frequent shipments to mitigate these fixed costs by placing larger orders. This finding is in line with \cite{kropf2014fixed}. 

After finding the shipping frequency responses to the financing costs, delivery time, and per-shipment fixed costs, I now investigate how the shipping frequency mechanism affects the financing demand to emphasize the relationship between financing demand and trade facilitation.

\subsubsection{Financing Demand}
The financing demand is an important factor for the exporting decision \citep{humphrey2009exporters,nagaraj2014financial}. Due to financial constraints in a developing country, the exporter can only borrow working capital for the order in process. Thus, the firm's demand for financial services is a simple measure of the amount of loan the firm holds at any given point in time in the entire inventory period. The financing demand of the exporter is $D=cx^*(.)$.

\textbf{Proposition 3.} \textit{a) All else unchanged, the financing demand of the exporter} \textit{(i) decreases for higher financing costs of the exporter: $\frac{\partial D}{\partial r}\leq 0$, (ii) decreases for higher financing costs of the importer: $\frac{\partial D}{\partial r_1}\leq 0$, (iii) decreases for longer delivery times: $\frac{\partial D}{\partial \Delta}\leq 0$}, and \textit{(iv) increases for higher per-shipment fixed costs: $\frac{\partial D}{\partial f}\geq 0$}.

\textit{b) All else unchanged, if per-shipment costs are high, then an increase in financing costs or delivery times results in a greater financing demand decrease than when per-shipment costs are low: $\frac{\partial^2 D}{\partial r \partial f} \leq 0$, $\frac{\partial^2 D}{\partial r_1 \partial f} \leq 0$, and $\frac{\partial^2 D}{\partial \Delta \partial f} \leq 0$.}

\textbf{Proof:} See Appendix-\hyperref[a1]{A}

The higher financing costs at both origin and destination reduce the order size, which reduces the exporter's financing demand. Similarly, the higher delivery times make order sizes smaller, which lowers the financing demand. From \textit{Proposition 2}, I can see that the higher per-shipment fixed costs increase order size. This requires a higher financing demand for the exporter. As less frequent shipments due to high per-shipment costs increase demand for working capital financing and bolster the severity of financial constraints, financial demand adjustment through shipping decisions is essential.

The exporter's financing demand is affected by the per-shipment costs because of the shipping frequency adjustment. At a lower per-shipment cost, the increase in shipping frequency reduces the exporter's financing demand, which is now less responsive to the higher financing costs and longer delivery times. This important finding indicates that the effect of an increase in the financing costs and delivery times on the financing demand is mitigated by the firm's ability to increase shipping frequency. Thus, if the developing countries can lower the fixed costs, they benefit more from lessened financing demand. 

\subsubsection{Procurement Cost}
In my model, an increase in the delivery time raise procurement costs adding to the literature that shows that delivery time is costly \citep{djankov2010trading,berman2012time,hummels2013time}. As the increase in procurement cost depends on the severity of the fixed trade barriers (e.g., the per-shipment costs), I now investigate how procurement costs are affected by the model variables, and, consequently, by the fixed costs.

\textbf{Proposition 4.}
\textit{a) All else unchanged, the procurement costs} \textit{(i) increase for higher interest rates in the exporter country: $\frac{\partial C}{\partial r } \geq 0$, (ii) increase for longer delivery times: $\frac{\partial C}{\partial \Delta } \geq 0$, (iii) increase for higher per-shipment costs: $\frac{\partial C}{\partial f } > 0$,} and \textit{(iv) decrease for higher interest rates in the importer country: $\frac{\partial C}{\partial r_1 } \leq 0$.}

\textit{b) All else unchanged, if per-shipment costs are high, then an increase in}  \textit{(i) exporter's interest rates or delivery time results in a greater cost increase than when per-shipment costs are low: $\frac{\partial^2 C}{\partial r \partial f} \geq 0$ and $\frac{\partial^2 C}{\partial \Delta \partial f} \geq 0$}, \textit{(ii) importer's interest rates result in a lower cost increase than when per-shipment costs are low: $\frac{\partial^2 C}{\partial r_1 \partial f} < 0$}.

\textbf{Proof:} See Appendix-\hyperref[a1]{A}

The procurement cost is a non-linear function of the optimized order size and relevant parameters, \\$C=f(x^*(.), \Delta, f, ...).$ Using this function, I find that the exporter financing costs, the delivery times or the per-shipment costs increase procurement costs. The simple intuition is that higher financing costs, longer delivery time, and high per-shipment costs require more working capital financing for a longer period at a higher per-shipment cost, raising the total financing costs and fixed costs. This, in turn, raises the procurement costs of an order. Thus, these variables increase the procurement cost of an extra order relative to the per-shipment costs of placing an additional shipment. The costs increase due to delivery time lag is in line with the time as fixed cost literature. However, the intuition is not simple for explaining why the importer's interest rate is reducing procurement costs. In this case, a higher interest rate increases the discounting of an order's immediate purchasing cost for the importer. Although the shipping frequency increases for a higher interest rate, the immediate purchases' accumulated present value is lower.

The increase in procurement costs due to longer delivery time or the financing costs depends on the per-shipment costs. For the higher per-shipment cost, the exporter financing costs and delivery time increase procurement costs at a higher rate than lower per-shipment costs. The intuition for this result is that the importer of a country with lower per-shipment fixed costs can increase shipping frequency. This mitigates the effect of the financing costs and delivery times on procurement costs. If firms can adjust shipping frequencies, greater financing costs and longer delivery times can be easily accommodated. However, for the importer financing costs, higher per-shipment costs, and delivery time decrease procurement costs at a higher rate due to the negative effect of importer interest rates on the procurement cost.

\subsection{Discussion}

The theoretical analysis gives some empirically testable predictions. From \textit{Proposition 2}, I find that the financing costs and the delivery time positively affect the shipping frequency, and the per-shipment costs negatively affect the shipping frequency. These predictions can be tested empirically with transaction-level trade data and financing data. Depending on the extent of financial development, the degree of capital market deepening, the severity of financial frictions, and the existence of capital control, the magnitude of the financing cost's effect on the shipping frequency may vary. Therefore, in the next section, I take testable predictions to a developing country's trade data to determine the effects on the shipping frequency. 

Depending on the data availability, \textit{Proposition 3} is testable if bank-firm matched data is available, and \textit{Proposition 4} is testable if exporter-importer paired data is available. The theoretical analysis changes if the per-shipment fixed costs are paid upfront. Here, the exporting firms are taking external funds immediately after receiving the importer's order to finance the per-shipment costs (see Appendix-\hyperref[a2]{A} for this alternative setup). In this case, the origin financing and the delivery times do not affect shipping decisions. Therefore, higher financing costs in the exporter country and longer delivery times do not affect the financing demand. Except for monitoring cost, this alternative finding is equivalent to the shipping size of the model of \cite{taylor1997competition}. Only empirical analysis can shed light on which modeling setup is appropriate for a country for its relevance with the data.

Using a partial equilibrium model to understand the financial demand adjustment through shipping frequency has limitations. I abstract away from other factors, such as price, demand effect, credit market, and the business cycle, important factors of shipping frequency decisions. I argue that in developing countries, the per-shipment fixed costs and the working capital financing costs are considerably higher than in developed economies. Furthermore, the exporter lacks bargaining power for pricing and shipping frequency decisions. Against this backdrop, the financing costs and the delivery times for emerging economies are essential to trade barrier issues for understanding the shipping frequency. However, the model's outcome will be undermined if the exporters use destination financing.

\section{Data and Estimation Strategy}
This section discusses the data and estimation strategies I will use in the estimation. 

\subsection{Firm-Level Trade Data}
Shipping frequency data is calculated from transaction-level export data from the Bangladesh Customs Authority, National Board of Revenue (NBR). The NBR collects customs data using ASYCUDA++, a computerized system designed by the United Nations Conference on Trade and Development (UNCTAD). I get this transaction-level data on export from the International Growth Center (IGC).\footnote{IGC facilitates research and provides policy advice in Africa and South Asia. They posted the Bangladeshi trade data on their website. The data can be downloaded from the following link: https://www.theigc.org/country/bangladesh/data/. For details about the accuracy of this data, see \cite{ahsan2017does}.} The data includes the date of export, the exporter's unique identification number, the freight on board export value, the export volume, Harmonized System eight-digit codes (HS8) of the product, the export destination, and the port of export. In the dataset, I have data from July 1, 2004, to December 31, 2013. In this study, I use annual data for all control variables. Thus, I eliminate data from July 1, 2004, to December 31, 2004. I lose 163,393 observations. Furthermore, observations with the missing firm identification number are eliminated. I also drop data from 2005 as the per-shipment fixed costs are missing for many countries. I lose 0.35 million observations. Finally, I trim the data at the extreme tail of the per-shipment export value to eliminate any potential outliers by eliminating the observations smaller than one percentile and larger than 99 percentiles. I lose 96,162 observations.\footnote{In the robustness checks, I show that including these observations does not affect the results.}

After data refinement, the final sample has 4.97 million transactions spanning eight years, where 18,327 firms export 4,234 HS8 products to 179 destinations. In the data, all monetary values are in constant 2010 Bangladeshi Currency, Taka (BDT). I convert these export values to match other macro variables using the average annual exchange rates against US dollars (USD). Table \ref{ch2t} summarizes the Bangladeshi export data. The total export volume is \$156.7 billion (in constant 2010 USD). The average shipping size is \$31,472, and the median shipment size is \$14,245. Most of the exports go to only a few developed trade partners, as reported in table \ref{ch2t1}. About 59.51 \% of exports shared by volume are going to the top 5 export destinations. The US covers 23.37 \% of export share by volume and 17.08\% share of shipments with 13.69 average shipments by the firm-product-year. The UK's export share by volume is 9.95 \%, but the shipment share is comparatively higher, 14.13\%. Thus, there is variation in shipping frequency relative to shipping volume across destinations, which I am interested in examining in the estimation sections.

I use the firm identifier to get the shipping frequency. A shipment is a single transaction at the port with a distinct Customs form. I aggregate all the shipments for a firm exporting an eight-digit HS code product to an exporting destination in a year. Thus, the shipping frequency variable captures all observations of firm-product-destination-year data combinations for 2006-2013. I find that there are 544,486 firm-good-year-destination observations. The largest firms make about two thousand shipments. However, the average shipping frequency for a firm is 8.92. Table \ref{ch2t2} provides the year-wise shipping frequency, the number of exporting firms, the number of new firms, and the number of firms with a single shipment. Column 2 of table \ref{ch2t2} reports considerable shipping frequency growth over the year. Therefore, in some specifications, I examine the effect of this growth by accounting for a year effect. The number of single shipment firms in column 4 shows that only 2,943 firms out of 18,327 have only single shipments throughout the study period. Thus, more than 80\% firms make multiple shipments of an HS8 product to a specific destination. I report high-frequency firms (more than 100 shipments) in column 5 of table \ref{ch2t2}, which shows the significant presence of high-frequency exporters. Finally, I calculate the per-shipment value by dividing the aggregated export value by shipping frequency.

\begin{figure}[H]
    \centering
    \renewcommand\thefigure{2}
    \includegraphics[width=150mm]{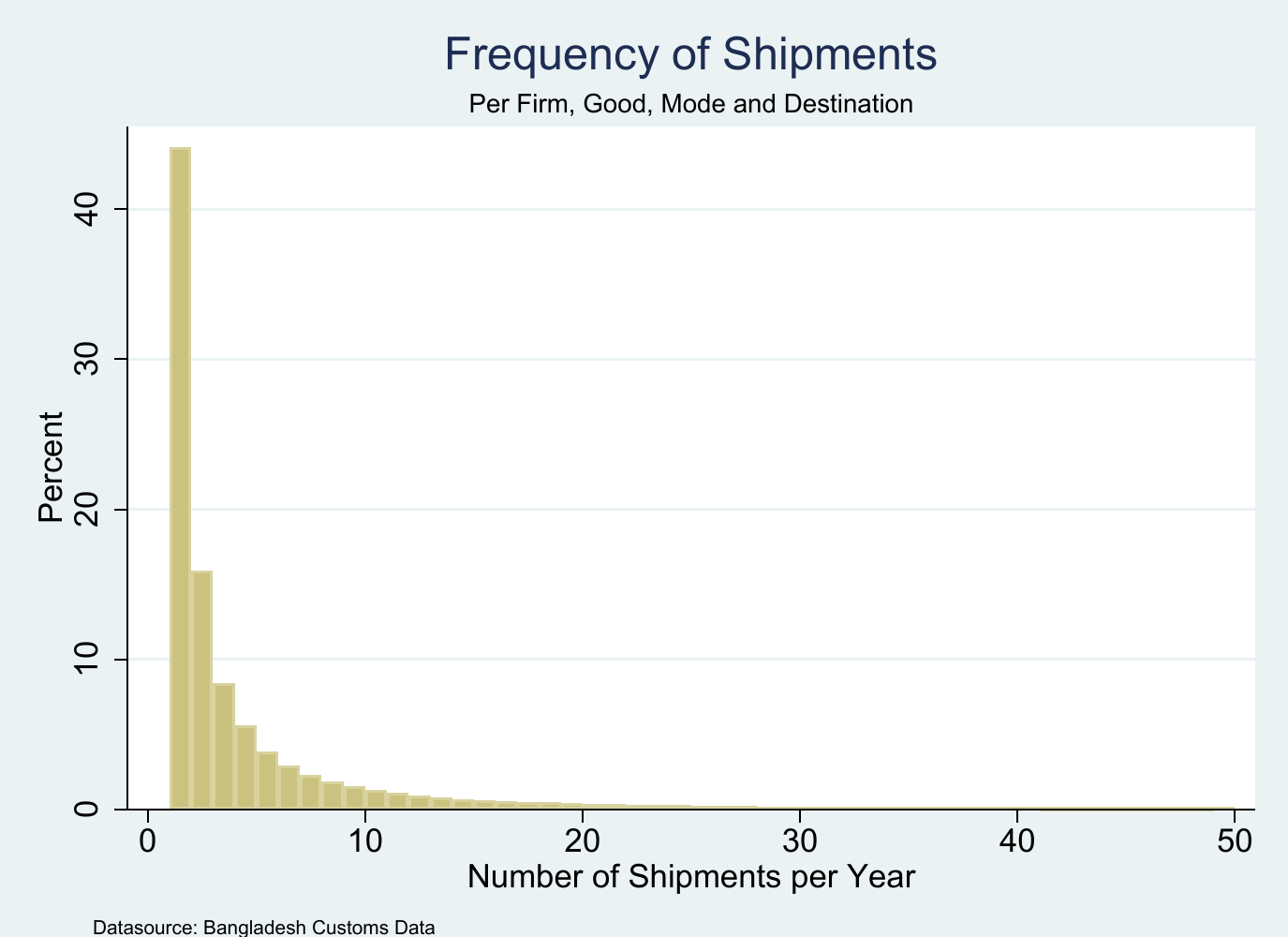}
    \captionsetup{justification=centering}
    \caption{Frequency of Shipments of Bangladeshi Exports, Data Combinations of Exporter-Product-Mode-Destination-Year for 2006-2013. The upper limit of shipping frequency is cut to 50 shipments.}
    \label{ch2f1}
\end{figure}

\begin{figure}[H]
    \centering
    \renewcommand\thefigure{3}
    \includegraphics[width=150mm]{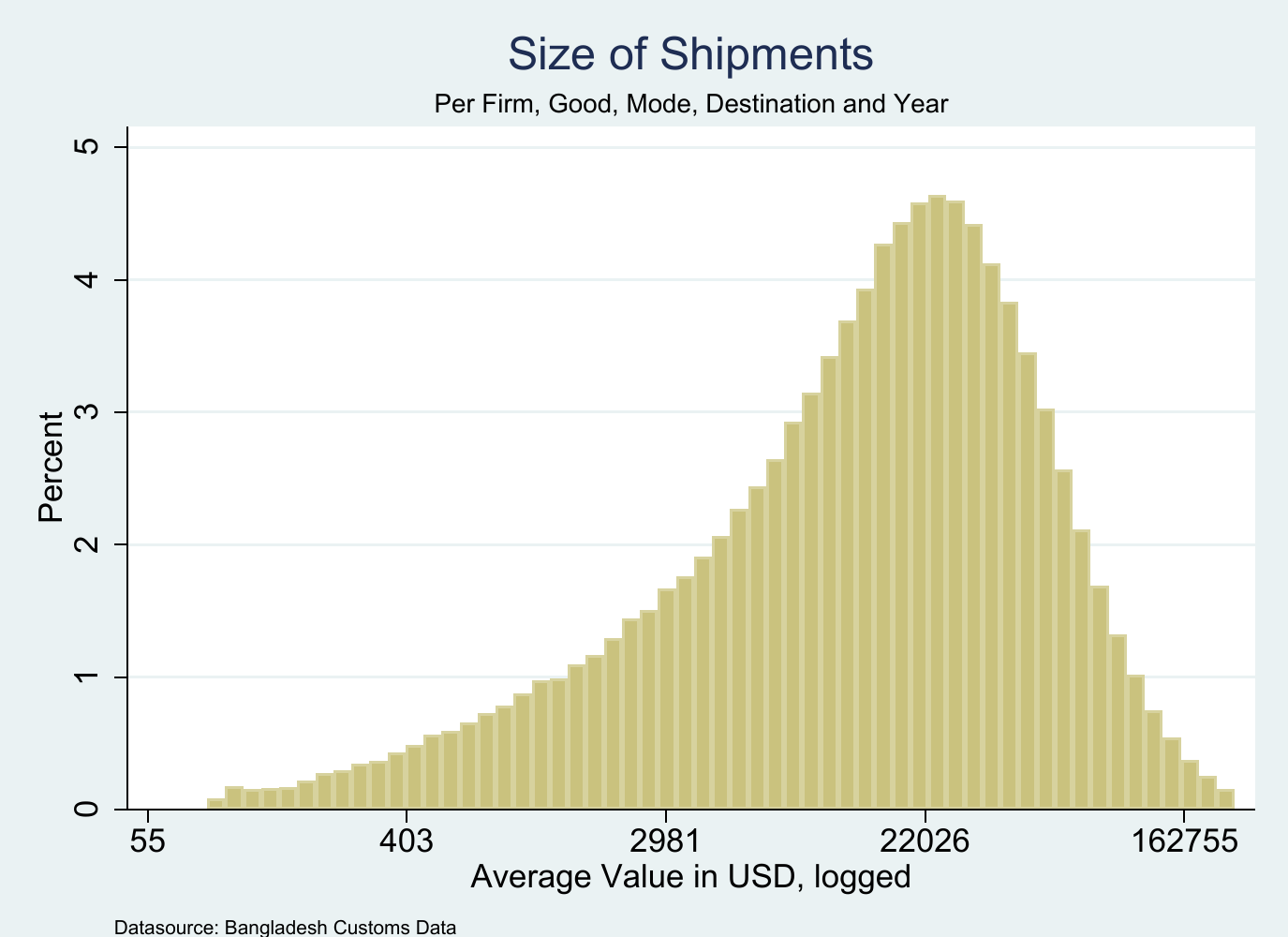}
    \captionsetup{justification=centering}
    \caption{Size of Shipments of Bangladeshi Exports in USD (logged), Data Combination of Exporter-Product-Mode-Destination-Year for 2006-2013. The upper limit of shipping size is cut at 3.27 million USD.}
    \label{ch2f2}
\end{figure}

Figure \ref{ch2f1} shows the shipping frequency for the exporter for 2006-2013. The figure confirms the stylized facts of international trade that shipments are infrequent and lumpy. Many single shipment firms contribute about 45\% of the shipments at the exporter-product-mode-destination-year observation level. The distribution of the size of the shipments is presented in figure \ref{ch2f2}. Excluding a few small and very few large-sized shipments, most shipping sizes are close to the median value of \$14,245. Figure \ref{ch2f2} shows that the distribution of shipping frequency is similar to a negative binomial distribution.

\subsection{Country-Level Data}
For the importer's financing cost, I get interest rates (lending rates) from the International Financial Statistics (IFS) of the International Monetary Fund (IMF). For the exporter's financing cost, I use scheduled bank average interest rates taken from the central bank of Bangladesh (Bangladesh Bank). 

I need additional data about the financial constraints of the destinations to deal with omitted variable bias. The debt to GDP ratio data comes from \cite{beck2009financial}. The debt-to-GDP ratio measures the level of financial development of an economy. I use a credit and bank confidence scores to consider the economy's financial frictions. A credit score measures accessibility to finance by considering the strength of credit reporting systems and the effectiveness of collateral and bankruptcy laws in facilitating interest. I get a credit score from the Doing Business Survey database of the World Bank for credit accessibility. For the cultural perception of the financial sector, I use confidence in bank measures from the World Values Survey (WVS) database \citep{inglehart2010wvs}. For alternative measures of interest rates, I use net interest rate margins taken from \cite{beck2009financial}. The net interest margin considers loss due to non-performing loans and captures the average ex-post markup on interest. It comes from the accounting value ratio of financial intermediaries' net interest revenues and their total earning assets. Therefore, the net interest margin captures the financial intermediaries' opportunity cost while the interest rates capture the importer's real financing cost if they borrow from the banking system. 

I take the per-shipment fixed costs from the Doing Business Survey (DBS) database of the World Bank. For per-shipment costs, DBS measures the costs of documentary compliance, border compliance, and domestic transport, excluding tariffs.

The bilateral distance (in kilometers, between most populated cities of a country-pair) data comes from the Center for International Prospective Studies (CEPII) Geodist dataset \citep{mayer2011notes}. I use sea-distance data from the CERDI sea-distance database \citep{bertoli2016cerdi} for alternative distance measurements. This database contains bilateral maritime distances between 227 countries and territories. The relevant port(s) for countries with the most shipping lines are considered for computing sea distance (landlocked countries are linked to the foreign port with the shortest road distance to its capital city). I calculate transport time from this database using an average speed of 13 knots.\footnote{The average container vessel speed comes from searates.com.} I add the import port delay with the transport time for the total time. Import port delay data is taken from the DBS database. I also use an alternative measure of the trade barrier, the Logistics Performance Index (LPI), which comes from the World Bank. 

Control variables such as GDP (in constant US dollars), GDP per capita (in constant US dollars), common religion (an index), island (a dummy), landlocked (a dummy), common legal origin (a dummy), and common colony (a dummy) are taken from the Center for International Prospective Studies (CEPII) Gravity dataset \citep{head2010erosion}.

The table \ref{ch2t3} includes all regression variables' summary statistics and data sources. 

\subsection{Estimation Strategy}

\cite{lewbel1992aggregation} discusses the presence of aggregation bias for log-linear panel data where mean-scaling is not ensured. This is likely true for shipping frequency data as the number of exporter shipping to a destination are correlated over the panel period. Furthermore, the more important source of bias is that in aggregate-level data, the number of shipments depends on the number of exporters and the number of times they ship. Thus, it is not clear if estimates capture the effect on the number of exporters or the number of shipments. To investigate this, I start with the country-level analysis, followed by product and firm-level analysis. 

For the country level, I estimate the following model of \cite{hornok2015administrative}:
\begin{equation}\label{ch2cols}
    \begin{multlined}
     ln\ N_{jt}=\alpha + \beta_1 ln\ f_{jt}+ \beta_2 ln\ distance_{j} + \theta X_{jt} + \lambda_{j}  +\xi_{t}+ \epsilon_{jt}
\end{multlined}
\end{equation}

The dependent variable is $ln\ N_{jt}$, the log of the number of shipments from Bangladesh to the country $j$ in year $t$. $ln\ f_{jt}$ is the log of the per-shipment fixed costs from Bangladesh to the importer country $j$ in year $t$. $ln\ distance_{j}$ is the log of the distance from Bangladesh to the export destination $j$. $X_{jt}$ contains control variables. For control variables, I use GDP and GDP per capita to control quantity and demand effects \citep{anderson2011gravity}. The dummies capture the demand and fixed costs for the island, landlocked, common legal origin, and common colony, and index for common religion \citep{head2014gravity}. Thus, $X_{jt}$ contains the log of $GDP_{jt}$, log of the per capita $GDP_{jt}$, island dummy, landlocked dummy, common colony dummy, common legal origin dummy, and common religion index. Although the common language is considered in \cite{hornok2015per} as a control variable, there is no trade partner except India, which speaks Bengali. Instead, I consider a common religion dummy for Bangladeshi trade. Bangladesh, an 89\% Muslim majority country, exports more religion-specific goods (e.g., Halal products) to Islamic countries. I include country fixed effect $\lambda_{j}$ to account for destination-specific unobservables. $\xi_{t}$ is year fixed effect. Finally, the error term is $\epsilon_{jt}$. The coefficients of interest $\beta_1$ capture the effects of the per-shipment fixed costs on the shipping frequency. Previous studies, as well as my theory, predict that $\beta_1<0$. A negative value of $\beta_1$ refers to reduced shipping frequency for higher per-shipment costs. According to previous studies, distance has a negative effect on the shipping frequency, which means $\beta_3<0$. 

The estimates from OLS estimation of log-linear models are biased \citep{silva2006log}. Furthermore, the shipping frequency is a count variable. Thus, the PPML estimation method is better suited to this empirical analysis. I use the following model\footnote{I use \textit{reghdfe} and \textit{ppmlhdfe} packages in Stata for OLS and PPML estimation. This can deal with high-dimensional fixed effects \citep{correia2017linear,ppmlhdfe,ExistenceGLM}.} 
\begin{equation}\label{ch2cppml}
    \begin{multlined}
     N_{jt}=exp\big[\alpha + \beta_1 ln\ f_{jt}+ \beta_2 ln\ distance_{j} + \theta X_{jt} + \lambda_{j}  +\xi_{t}\big] \times \epsilon_{jt}
\end{multlined}
\end{equation}

Being a developing country located in a specific geographic location, distance is non-linearly related to trade flow and shipping frequency. Especially, the high export share with the EU (biggest export destination) and the US (second biggest export destination) create the non-linearity between shipping frequency and distance. Investigating shipping data, I find three splines provide the most reasonable fix for distance non-linearity. The first spline (spline1) includes only neighbor countries inside 3970 kilometers radius, the second spline (spline2) includes European countries and other destinations beyond neighbor countries inside 9283 kilometers radius, and the third spline (spline3) includes the US and the rest of the destinations. Thus, I modify model \ref{ch2cppml} for distance fix by introducing two dummies for three distance splines. The modified model is as follows:
\begin{equation}\label{ch2cppmln}
    \begin{multlined}
     N_{jt}=exp\big[\alpha +spline1+spline2+ \beta_1 ln\ f_{jt}+ \beta_2 ln\ distance_{j} + \beta_3 spline1 \\ \times ln\ distance_{j} + \beta_4 spline2 \times ln\ distance_{j} + \theta X_{jt} + \lambda_{j}  +\xi_{t}\big] \times \epsilon_{jt}
\end{multlined}
\end{equation}

Next, I focus on the product-level analysis and estimate the model of \cite{hornok2015per}:
\begin{equation}\label{ch2pols}
    \begin{multlined}
     lnN_{gmjt}=\alpha + \beta_1 ln\ f_{jt}+ \beta_2 ln\ distance_{j} + \theta X_{jt} + \eta_{g}+ \phi_{m}  +\xi_{t}+ \epsilon_{gmjt}
\end{multlined}
\end{equation}

Where $N_{gmjt}$ is the number of the shipments from a firm in Bangladesh exporting product $g$ by transport mode $m$ to country $j$ in year $t$. I include products fixed effect $\eta_{g}$ to account for product-specific unobservables. Transport mode-specific shocks are absorbed by $\phi_{m}$. $\epsilon_{gmjt}$ is the error term. Other notations are similar to the previous model. 

My preferred specification comes from \ref{ch2cppmln} with product-level modification
\begin{equation}\label{ch2pppmln}
    \begin{multlined}
     N_{gmjt}=exp\big[\alpha +spline1+spline2+ \beta_1 ln\ f_{jt}+ \beta_2 ln\ distance_{j} +  \beta_3 spline1 \\ \times ln\ distance_{j} + \beta_4 spline2 \times ln\ distance_{j} + \theta X_{jt} + \eta_{g}+ \phi_{m} +\xi_{t}\big] \times \epsilon_{gmjt}
\end{multlined}
\end{equation}

To overcome the aggregation bias, now, I consider firm-level analysis by modifying model \ref{ch2pppmln} 
\begin{equation}\label{ch2fppmln}
    \begin{multlined}
     N_{igmjt}=exp\big[\alpha +spline1+spline2+ \beta_1 ln\ f_{jt}+ \beta_2 ln\ distance_{j} +  \beta_3 spline1 \\ \times ln\ distance_{j} + \beta_4 spline2 \times ln\ distance_{j} + \theta X_{jt} + \lambda_{i}+ \eta_{g}+ \phi_{m} +\xi_{t}\big] \times \epsilon_{igmjt}
\end{multlined}
\end{equation}

Where $N_{igmjt}$ is the number of the shipments from a firm $i$ in Bangladesh exporting product $g$ by transport mode $m$ to country $j$ in year $t$. $\lambda_{i}$ is the firm fixed effect, which absorbs firm heterogeneity. $\epsilon_{igmjt}$ is the error term. 

The model of procurement of the previous section shows that higher interest rates and longer delivery times increase financing costs, reduce shipping size and increase shipping frequency. The intuition for these predictions is that higher interest rates in the exporter country increase the exporter's costs for borrowing working capital. This increases the importer's immediate procurement cost. To mitigate this cost increase, the importer chooses more frequent small-sized shipments. Furthermore, higher discounting due to higher interest rates in the importing country increases the shipping frequency by moving more small shipments into the future. By adding financing costs to the model \ref{ch2fppmln}, I estimate the following model
\begin{equation}\label{ch2fppmlnf}
\centering
    \begin{multlined}
     N_{igmjt}=exp\big[\alpha +spline1+spline2+ \beta_1 ln\ f_{jt}+ \beta_2 ln\ distance_{j} + \beta_3 spline1 \\ \times ln\ distance_{j} + \beta_4 spline2 \times ln\ distance_{j}+ \beta_5 r^{exp}_{t}+ \beta_6 r^{imp}_{jt}+\beta_7 r^{exp}_{t}\times r^{imp}_{jt}\\+ \theta X_{jt} + \lambda_{i}+ \eta_{g}+ \phi_{m} +\xi_{t}\big] \times \epsilon_{igmjt}
    \end{multlined}
\end{equation}

Where the exporter's interest rates are $r^{exp}_{t}$ in year $t$, in percentage, the importer's interest rate is $r^{imp}_{jt}$ of the importer country $j$ in year $t$, expressed in percentage. The interaction effect of the exporter's and importers interest rates is captured by $\beta_7$. Higher financing costs in origin and destination should increase shipping frequency; I expect positive coefficients for $\beta_5-\beta_7$. For the PPML estimation method, the dependent variable is in level, whereas the dependent variable is in logs for the OLS estimator. $\beta_5-\beta_7$ in the model capture semi-elasticity as the interest rates of the exporters and the importers (independent variables) are in levels. 

It is assumed that the exporting firms face their domestic interest rates as the funding costs for trade finance of working capital. Similarly, the importing firms consider discounting based on their domestic interest rates. Thus, firm-specific characteristics do not affect the financing cost. In most specifications, exporter-product-mode-year specific unobservables (e.g., common shocks to the firm-product-mode-year observation) are absorbed through the firm by product by mode by year fixed effects to capture only interest rate variations across the destinations. The standard errors are robust and clustered with the exporter group clusters. As the disturbances are correlated within exporters, robust clustered standard errors can account for the correlation of the exporters' observations. 

Based on the empirical models above, I will consider several types of fixed effects for absorbing relevant heterogeneity. I will also conduct additional empirical exercises focusing on other measures of financing costs. Furthermore, I will provide several robustness checks considering the additional distance and time measures and sample restrictions.

\section{Results}
I run empirical analysis in three steps. First, I show that the findings of the higher per-shipment costs decrease shipping frequency remains valid for a developing country. Second, I show that financing costs affect the shipping frequency. Finally, I check whether the main specification's findings remain consistent with alternative measures of financing costs.

\subsection{Per-Shipment Costs and Shipping Frequency}
For the country-level aggregation, the estimation results for regressions of model \ref{ch2cols} are presented in table \ref{ch2tbase}. In column 1, I present the OLS regression results with destination-fixed effects. Standard errors are robust and clustered with destination group clusters. The coefficient for the per-shipment cost (including fixed import costs) is insignificant. Considering the total per-shipment cost (includes fixed export and fixed import costs), \cite{hornok2015administrative} finds -0.451, with a $p$-value of 0.24. Considering only year fixed effects, I get a negative effect of per-shipment fixed cost on shipping frequency with an elasticity of -0.678, significant at a 1\% level. To deal with heteroscedasticity biases of OLS estimation of the log-linear model, I consider the PPML estimation method for model \ref{ch2cppml} in column 3. The per-shipment cost coefficient becomes non-significant. When I include distance splines to fix distance non-linearity and estimate model \ref{ch2cppmln}, the elasticity is smaller and significant at a 5\% level, as reported in column 4. 

Now, I consider the product-level aggregation model \ref{ch2pppmln} and report the results in columns 5-7 of table \ref{ch2tbase}. In column 5, I present the OLS regression results with eight-digit HS product by transport mode by year fixed effects. Standard errors are robust and clustered with exporter group clusters. The per-shipment cost coefficient is very small and non-significant. I find a much stronger response for per-shipment cost in the PPML estimation method than in the OLS method, as reported in column 6. By including distance splines, the coefficient for per-shipment costs is even stronger. I find that for a 10\% increase in the per-shipment costs; the shipping frequency reduces by $(1.1^{-0.406}-1)\times 100=3.95$\%. This is a larger shipping frequency response for the per-shipment cost for a developing country (e.g., Bangladesh) than \cite{hornok2015per} findings of developed countries (2.53\% for the US and 0.99\% for Spain). Considering PPML and OLS estimation approaches, the advantage of PPML is that it is robust concerning heteroscedasticity but suffers from inconsistency with many fixed effects due to the incidental parameters problem.

To check whether there is aggregation bias, I now regress model \ref{ch2fppmln} for exporter-level data. In column 8 of table \ref{ch2tbase}, I present the PPML regression results with same-year fixed effects considered in the country-level analysis. The shipping frequency response of per-shipment costs is less than half that in column 4. The elasticity is -0.156. Thus, a 10\% increase in the per-shipment costs reduces the shipping frequency by $(1.1^{-0.156}-1)\times 100=1.49$\%. Column 9 shows exporter-level analysis with the same fixed effects of product-level. The export level's per-shipment elasticity is -0.227, much lower than the product-level elasticity of -0.406. Comparing columns 4 and 7 with columns 8 and 9, I find that the per-shipment cost estimates are twice as large with similar fixed effects in the country and exporter levels. This shows evidence of aggregation bias. At a more aggregated level, the per-shipment cost elasticity is higher because of both the number of shipments and the number of exporter effects. In the exporter-level data, the per-shipment cost elasticity comes from only the number of shipment effects. 

Now, I focus on the effects of distance on shipping frequency. The distance coefficient is negative and significant at the country level, as reported in column 2 of table \ref{ch2tbase}. However, the elasticity is much smaller than that reported in \cite{hornok2015administrative}. At the product level, the coefficient for distance is not significant unless the distance fix is considered. Column 7 shows that the distance coefficient is negative for spline1 (elasticity, -2.268+1.018=-1.25) and spline3 (elasticity, -2.268) but positive for spline2 (elasticity, -2.268+3.713=1.445). Thus, for Bangladeshi trade, the effect of the distance on shipping frequency depends on the region, mostly negative and in line with \cite{hornok2015administrative,hornok2015per} for non-EU regions. Overall, distance has a non-linear effect, and accounting for that is important to determine the effect of per-shipment costs on shipping frequencies. 

The size and income level of the export destinations capture the demand effect and determine the shipping frequency. Here, I investigate GDP and GDP per capita. At the country level, considering the destination fixed effects, the effect of GDP is 3.7 times higher than that found in \cite{hornok2015administrative}. GDP and GDP per capita positively affect shipping frequency across all other specifications, as shown in columns 2-9 of table \ref{ch2tbase}. At the product level, these elasticities are much larger in Bangladesh than in developed countries. Overall, the effects of GDP and GDP per capita increase with a higher level of aggregation.

I also estimate and compare other regressors. In column 9 (exporter-level specification), the island dummy's coefficients are flipped from the previous columns. A very small and non-significant island coefficient is different from the finding of \cite{hornok2015per}, due to very concentrated export to the non-island destinations. The landlocked dummy coefficients are significant and negative, capturing a high trade barrier effect on the shipping frequency. The coefficients for common religion, common legal origins, and common colony are positive, capturing a low trade barrier effect on the shipping frequency. Except for the island dummy, these control variable coefficients are similar in sign and significance, larger in magnitude than the findings of \cite{hornok2015per}. Except for the island and landlocked dummy, these control variables' magnitudes are higher at higher-level of aggregation.

\subsection{Omitted Variable Bias and Financing Costs}

Now I deal with omitted variable bias by adding narrower fixed effects. As the shipping decisions are made at the firm level, absorbing the variations at a broader level might weaken the findings. Therefore, I verify whether the results remain robust for a stricter specification. I start with extensive fixed effects and narrow the fixed effects to check if the estimates remain consistent. The results are presented in table \ref{ch2fe}. The per-shipment cost's effect on shipping frequency gets stronger from the broader to stricter fixed effects. The elasticity ranges from -0.179 for exporter fixed effects in column 1 to -0.331 for exporter-HS4-mode-year fixed effects in column 6.\footnote{For firm by-product by year fixed effects, lots of singleton observations are dropped. For reducing these lost observations and from a shipment perspective, goods classification is best captured by HS4 instead of HS8. For example, consider the HS8 code for cotton T-shirts, 61099000, a textile and apparel industry good. The chapter heading on apparel and clothing knitted is 61. The next two digits, 09, are for the sub-heading T-shirt. The next two digits, 90, are for materials sub-divisions. Finally, the last two digits, 00, are for cotton.} The coefficients for GDP and GDP per capita hold for both broader and stricter fixed effects. Other control variables have robust responses too. For example, Island and colony coefficients are more significant for narrower fixed effects. Overall, absorbing more omitted variables with stricter fixed effects increases the elasticity of per-shipment fixed costs. Column 6 is my preferred specification as it absorbs variations for any exporter shipping HS four-digit product by a particular shipping mode in a specific year. In the preferred specification, a 10\% increase in the per-shipment costs reduces the shipping frequency by $(1.1^{-0.331}-1)\times 100=3.21$\%.

Next, I check the theoretical prediction regarding the effects of financing costs on shipping frequency to deal with omitted variable bias. Table \ref{ch2fin} reports regressions results for model \ref{ch2fppmlnf}. Introducing the importer financing costs (interest rates) lowers the per-shipment cost effect on the shipping frequency; the coefficient changes from -0.227 to -0.196. This confirms that per-shipment cost effects are overestimated, and financing costs can capture some per-shipment effects. Column 1 also shows that the importer's interest rate has a negative effect on the shipping frequency. As the coefficient captures semi-elasticity, for a 1\% higher interest rates in the importer country, the shipping frequency goes down by $(\mathrm{e}^{0.004}-1) \times 100=0.40$\%, which contradicts the prediction. One explanation for these contradictory findings may come from the deviation from the theoretical assumptions that the importer has most of the bargaining power and that the exporter is financially constraint. Although these conditions are more likely met for exporters in Bangladesh, I do not have any way to verify this in the data. Another potential reason could be the theoretical model's omission of some trade costs. For example, the Generalized Scheme of Preferences (GSP) for the Bangladeshi exporter in the US (continued until the study period, 2013) and the EU reduces trade costs \citep{ahmed2009sustaining}. Allowing year variation, column 2 shows that the exporter's interest rate positively affects the shipping frequency, supporting the theoretical prediction.

Next, I check the interaction between the exporter's and the importer's interest rate in column 3 of table \ref{ch2fin}. The coefficient is positive and significant at a 1\% level, which implies that shipping frequency increases more for higher interest rates both in origin and destination. However, the interaction effects are very small in magnitude. In column 4, I use a narrower exporter by eight-digit HS product by mode by year fixed effects. The per-shipment cost elasticity is much higher than the elasticity of product and country-level analysis. The interaction term is also stronger. From now on, I will consider this specification as the main specification and the findings of column 4 as the main findings. Thus, by addressing all the biases, I find that for a 10\% increase in the per-shipment costs, the shipping frequency decreases by $(1.1^{-0.335}-1)\times 100=3.24$\%.

In summary, columns 1-4 of table \ref{ch2fin} show that omitting financing costs may lead to minimal overestimation of per-shipment cost for allowing exporter heterogeneity to a severe underestimation of per-shipment cost for absorbing exporter heterogeneity. The effect of per-shipment costs increases when I add the financing costs of the exporter and importer and their interaction, as reported in column 4. Thus, my preferred specification shows that models underestimate the effects of per-shipment cost on shipping frequency by ignoring financing costs and other unobserved variables. 

The shipping frequency response also depends on how fast transport is done. Longer delivery time in ocean transport requires more financing resulting in more frequent shipments. Furthermore, time-sensitive goods are prevalent in air shipments, which mute the financing costs effect on shipping frequency because of the urgency in the supply chain \citep{hummels2007transportation,hummels2013time}. Thus, considering transport mode can reveal whether the interest rates have a stronger shipping frequency response for lengthy ocean transport. I consider transport mode effects in columns 5-7 of table \ref{ch2fin}. Column 5 shows that the financing costs effects are negligible for air freight goods. The shipping frequency response for the per-shipment cost is stronger in ocean shipment than in the air shipment mode, as reported in column 6. Moreover, the importer interest rates have a higher negative effect on ocean mode. The coefficients for other control variables remain similar to the main results. With only 4643 observations, the land freight mode shows that the per-shipment cost elasticity is stronger than the main results. The land transport mode also shows the positive effects of exporter financing costs, matching the theoretical predictions. As there are only three destinations for land transports of Bangladesh \citep{kathuria2018glass}, time-invariant variables drop out of the regression. Overall, ocean freight goods' shipping frequency responds more to the per-shipment and financing costs than other transport modes.

\subsection{Additional Measures of Financing Costs}
Facing a few data points for the exporter's annual interest rates, I now focus on the destination's financing costs. To determine the full impact of the financing costs on the shipping frequency, I consider other influencing variables of financing costs such as credit accessibility, bank confidence indicator, and the export destinations' credit-GDP ratio in the main specification as only interest rates might not capture the financial constraints of the firm. Columns 1-3 of table \ref{ch2finalt} report this result. Higher credit accessibility facilitates trade by increasing trade volume \citep{bremus2018reduced}. This explains a positive effect on shipping frequency for greater accessibility to credit, as reported in column 1. Bank confidence indicators capture the cultural dimension of financing costs. Financial literacy and cultural perceptions are key factors in the financial environment \citep{dutta2012culture}.\footnote{I use bank confidence indicators from the WVS database, which encodes lower confidence in the banking system as a higher indicator value. The indicator is calculated by aggregating the country-level average of the values chosen by all survey respondents of a country. Low confidence shows perceptional mistrust about financial services and weak financial development, increasing the financial costs of any credit contract. Consequently, this increases the shipping frequency.} As expected, weaker confidence in the banking system has a strong positive significant effect on the shipping frequency, as reported in column 2. The literature uses the credit-GDP ratio to measure the credit constraints \citep{manova2012credit,fauceglia2015credit}. Column 3 shows a minimal negative effect of the credit deepening on the shipping frequency. The finding aligns with the prediction that low financing costs lead to less frequent shipments. Across these three specifications, the interest rate coefficients are negative and significant. Furthermore, per-shipment cost effects remain the same by controlling for the credit-GDP ratio, increasing bank confidence, and decreasing by controlling for access to credit.

Now, I use the alternative measure of financing costs. The net interest margin is a common metric in finance studies \citep{beck2009financial,yuksel2017influencing}. I find a significant negative coefficient of interest rate interaction in column 4 of table \ref{ch2finalt}. One reason for this anomalous result could be that the net interest rate may not capture the importing firm's actual financing cost of borrowing from the banking channel; instead, it captures the financial intermediary's opportunity cost for assets. Across columns, other coefficients are roughly similar in sign and magnitude to the main results. In summary, controlling for additional financing measures does not change the main financing cost's effect on shipping frequency but changes the magnitude of per-shipment cost's effect on shipping frequency.

\section{Robustness Check}
I run several robustness checks. First, I investigate whether the findings remain consistent if I control for additional distance and time measures. Sea distance, transport time based on sea distance, total transport time with port delay, and logistics performance index are used for additional control. Table \ref{ch2time} reports these results. As sea distance is relevant for only ocean-transported goods, columns 1-3 only consider sea transport observations. In columns 1 and 2, I consider sea distance and transport time based on sea distance, respectively. The shipping frequency response for per-shipment costs is less than half of the main results. However, the per-shipment cost elasticity is closer to the main findings for the total transport time, as reported in column 3. LPI captures the supply chain and logistical barriers for exports \citep{marti2014importance}. In column 4, I find a positive effect of LPI on shipping frequency, which means lower logistical costs for importing increase shipping frequency. 

Second, I use different group clusters for calculating standard errors to see if the regression statistics are affected by different clusters. Table \ref{ch2se} reports the results for the preferred specification with the exporter by eight-digit HS product by mode fixed effects. I narrow down to exporter by industry (HS2) by destination by year clusters, starting with bigger clusters of export destinations. The columns show that the coefficient's standard errors get smaller with the larger numbers of clusters. For example, for the coefficient of per-shipment cost, the standard error is 0.085 for exporter clustering and 0.019 for exporter by industry (HS2) by destination-by-year clustering. However, different clustering does not change the basic statistical conclusion of the analysis.

Third, I consider various subsamples based on data characteristics. I check whether the main findings remain consistent for the subsamples based on relevant data restrictions. Table \ref{ch2sub} reports this result. All columns show regression for estimation model \ref{ch2fppmlnf}. In column 1, I consider the observations with an export value of more than 2500 USD to compare with the previous studies \citep{kropf2014fixed,hornok2015per}. To exclude data anomaly from 2007 (the total shipment number and export volume decreased in Bangladesh due to political instability) and global recession effects, I only consider data from 2010 to 2013 in column 2. Column 3 drops all single shipment firms. In column 4, I exclude all high-frequency firms with more than 100 shipments to a product-destination-year triplet. Column 5 reports result for the full sample, including smaller than 1\% and larger than 99\% per-shipment value. Excluding high-frequency exporters reduces the elasticity by half, as reported in column 4. Across all other columns, the results are robust to these sample restrictions. 

Finally, to investigate whether the per-shipment and financing costs effect are robust on the shipping size and export volume (by value or by weight), I regress these variables on the explanatory variables and report the result in the Appendix-\hyperref[a1]{B}. Except for the effect of the per-shipment cost on sipping size, the results align with \cite{hornok2015per}.

\section{Conclusion}

Financial constraints and shipping frequency are interrelated in international trade. I develop a simple export-import procurement model to capture this, incorporating delivery time, financing costs, and per-shipment fixed costs. The model shows that longer delivery times and higher financing costs increase shipping frequency because of financial constraints. The existing shipping frequency or financial constraints literature does not discuss this shipping frequency adjustment. 

The shipping frequency and trade finance aspects capture the significance of the time costs of trade. The procurement cost is increased for longer delivery time and higher financing costs, more so at higher per-shipment costs than lower per-shipment costs. The conjunction of the financing cost, the per-shipment cost, and the delivery times provide key insights about the procurement costs, which are not explored in the delivery time literature or time trade barrier literature.  

The model also predicts that higher per-shipment fixed costs decrease the shipping frequency in line with previous studies. This, in turn, shows that the exporter's financing demand decreases with the lower per-shipment fixed costs. From a policy perspective, this shows that if trade facilitation policy lowers per-shipment fixed costs, then that not only lowers transaction costs but also relaxes financial constraints. 

As understanding the shipping frequency margin is emphasized due to the nature of lumpy trade and transmission of various macro shocks to the exporter, in this empirical exercise, I find that more aggregated data provides over-estimated elasticity for per-shipment costs. I also find that the OLS estimation of per-shipment costs of log-linear models is underestimated due to heteroscedasticity bias. Moreover, ignoring financing costs yields underestimated per-shipment costs effects on shipping frequency because of omitted variable bias. By addressing these biases, I find that statistically, for 10\% higher per-shipment costs, the shipping frequency decreases by 3.24\%. I also find that higher financing costs in origin and destination lead to more frequent shipments. These findings remain consistent for various alternative financial constraints, distance, or time measures.

The higher elasticity of shipping frequency for per-shipment costs in the developing country indicates that the per-shipment fixed costs reduction policies will benefit from lower financing demand and conventional advantages of low trade barriers. For developing countries (e.g., Bangladesh), if trade policies can reduce per-shipment cost and facilitate financing aid \citep{auboin2007boosting}, it may have more substantial trade-enhancing effects through shipping frequency adjustment. Although the magnitude of shipping frequency is useful to determine the impact of reducing trade costs, this paper's policy implications are incomplete due to limited data points of the exporter's actual financing costs. For proper identification, firm-level financing costs are needed. One can use the bank-firm merged financing costs data to overcome this problem \citep{paravisini2014dissecting,fan2015trade}. As lowering trade costs and providing financial services are vital components of trade aid, these policy evaluations could be one avenue of future research.

\pagebreak

\bibliographystyle{apalike} 
\bibliography{main}
\pagebreak

\vspace{-1.5em}
\footnotesize{Note: The bottom arrow represents order time between two consecutive orders. The top arrow represents the delivery time between the order initiated and the goods delivered.}

\pagebreak

\pagebreak

\pagebreak

\begin{table}[H]
    \centering
    \singlespacing
    \scriptsize
    \renewcommand\thetable{1}
    \caption{Exports of Bangladesh (2006-2013)}
    \begin{tabular}{l l }
    \toprule
       Total exports volume (billion US\$) & 156.7 \\
       Number of exporters &  18,327\\
       Number of products & 4,234\\
       Number of destinations & 179\\
       Number of shipments &  4,979,958\\
       Average shipment value (US\$) & 31,472 \\
       Median shipment value (US\$) & 14,245 \\
    \bottomrule
    \bottomrule
    \end{tabular}
    \label{ch2t}
\end{table}

\vspace{1cm}

\begin{table}[H]
    \centering
    \singlespacing
    \scriptsize
    \renewcommand\thetable{2}
    \caption{Top Export Destinations}
    \begin{tabular}{lcc} 
    \toprule
       Destination   & Export share (\%) & Export share (\%)  \\
         & by volume & by shipping frequency \\
         \midrule
         United States & 23.37 & 17.08 \\
         Germany & 15.45 & 10.81 \\
        United Kingdom & 9.95 & 11.72\\
        France & 6.28 & 6.44\\
        Spain & 4.46 & 3.43 \\
        \bottomrule
        \bottomrule
    \end{tabular}
    
    \label{ch2t1}
\end{table}
\vspace{1cm}

\begin{table}[H]
    \centering
    \singlespacing
    \scriptsize
    \renewcommand\thetable{3}
    \caption{Descriptive Data of Shipping Frequency}
    \begin{tabular}{c c c c c}
        \toprule
        
        Year & Shipping frequency & Exporting firms & Single shipment & High frequency \\
         
          & (million) & & firms & firms (shipment$\geq$100) \\
      \midrule
         
        2006 & 0.46 & 5,789 &  752  & 602  \\
        
        2007 & 0.35 & 5,828 &  756 & 430 \\
        
        2008 & 0.49 & 6,579 &  926 & 673 \\
        
        2009 & 0.52 & 6,678 & 962 & 716 \\
        
        2010 & 0.64 & 6,947 &  1,027 & 944 \\
        
        2011 & 0.68 & 7,219 & 1,074 & 996 \\
        
        2012 & 0.85 & 7,622 &  1,142 & 1,278 \\
        
        2013 & 0.98 & 9,374 &  1,263 & 1,287 \\
        
      \midrule
        Total & 4.98 & 18,327 &  2,943 & 7,298 \\
      \bottomrule
      \bottomrule
    \end{tabular}
    \label{ch2t2}
\end{table}

\pagebreak

\begin{table}[H]
    \centering
    \singlespacing
    \scriptsize
    \renewcommand\thetable{4}
    \caption{Summary of Regression Variables}
    \begin{tabular}{l c c c c l}
        \toprule
        Variable & Mean & Std. Dev. & Min & Max & Data Sources\\
        \midrule
       \textbf{Export}   &  &  &  &  &  \textbf{Bangladesh Customs}\\
       Ln shipping frequency  & 1.05 & 1.23 & 0 & 7.57 & \\
       Ln per-shipment value  & 9.42 & 1.58 & -1.73 & 20.33 &\\
       Ln export value   & 10.48 & 2.13 & -1.73 & 20.33 &\\
       Ln export weight  & 8.09 & 2.29 & -4.61 & 19.79 &\\
       
       &  &  &  &  &\\
       \textbf{Distance and time}   &  &  &  &  &\\
       Ln per-shipment costs  & 7.08 & 0.39 & 5.90 & 9.89 & Doing Business Survey database (WB)\\
       Time to import (days) & 9.35 & 5.61 & 4 & 117 & Doing Business Survey database (WB)\\
       Port to port transport time (days) & 22.67 & 6.69 & 0 & 37 & searates.com\\
       Ln distance & 8.88 & 0.51 & 6.06 & 9.81 & CEPII distance dataset\\
       Ln Sea-distance & 9.45 & 0.38 & 7.81 & 9.98 & CERDI  sea-distance  database\\
       Logistics performance index & 2.87 & 0.57 & 1.34 & 4.11 & LPI database (WB) \\
       &  &  &  &  &\\
      
       \textbf{Financing costs (\%)}   &  &  &  &  &\\
       Exporter's interest rates    & 12.69 & 0.93 & 11.30 & 13.77 & Bangladesh Bank\\
       Importer's interest rates   & 5.15 & 3.52 & 0.5 & 58.98 & International Financial Statistics (IMF)\\
       Exporter's net interest margin   & 4.29 & 0.99 & 2.77 & 5.57 & \cite{beck2009financial}\\
       
       Importer's net interest margin    & 2.30 & 1.45 & 0.12 & 20.48 &\cite{beck2009financial}\\
       Debt-GDP ratio & 123.04 & 50.21 & 2.17 & 906.38 &\cite{beck2009financial}\\
       
        &  &  &  &  &\\
       \textbf{Control variables}  &  &  &  &  &  \textbf{CEPII gravity dataset}\\
       Ln GDP & 27.87 & 1.49 & 18.46 & 30.45 &\\
       Ln GDP per capita & 10.23 & 0.96 & 5.09 & 11.63 &\\
       Island & 0.13 & 0.34 & 0 & 1 &\\
       Landlocked & 0.03 & 0.17 & 0 & 1 &\\
       Common religion & 0.09 & 0.23 & 0 & 0.86 &\\
       Common colony& 0.07 & 0.26 & 0 & 1 &\\
      \midrule
       N = &  & & &544,486 &\\
        \bottomrule
        \bottomrule
    \end{tabular}
    \label{ch2t3}
\end{table}

\pagebreak
\begin{landscape}
\begin{table}[H]
  \centering
  \tiny
  \singlespacing
  \renewcommand\thetable{5}
  \caption{Per-Shipment Cost and Shipping Frequency}
    \begin{tabular}{lccccccccc}
    \toprule
     & \multicolumn{4}{c|}{Country } & \multicolumn{3}{c|}{Product } & \multicolumn{2}{c}{Exporter}\\
     \cmidrule(lr){2-5} \cmidrule(lr){6-8} \cmidrule(lr){9-10}
          & (1)   & (2)   & (3)   & (4)   & (5)   & (6)   & (7)   & (8)   & (9) \\
    
    \midrule

    Ln per-shipment cost& -0.062 & -0.678*** & -0.174 & -0.359* & -0.054 & -0.307* & -0.406** & -0.156*** & -0.227*** \\
          & [0.183] & [0.185] & [0.217] & [0.187] & [0.078] & [0.179] & [0.202] & [0.030] & [0.030] \\

    Ln distance &       & -0.411*** & -0.456** & -1.991* & -0.098 & -0.048 & -2.268*** & 0.463*** & 0.084 \\
          &       & [0.140] & [0.225] & [1.159] & [0.076] & [0.156] & [0.715] & [0.157] & [0.161] \\
    Spline1 &       &       &       & -1.241 &       &       & -11.733* & 6.860*** & 3.467*** \\
          &       &       &       & [10.555] &       &       & [6.834] & [1.217] & [1.262] \\
    Spline2 &       &       &       & -31.388*** &       &       & -34.290*** & 1.165 & -3.909** \\
          &       &       &       & [11.421] &       &       & [8.385] & [1.691] & [1.644] \\
    
    Spline1$\times$Ln distance &       &       &       & -0.186 &       &       & 1.018 & -0.762*** & -0.423*** \\
          &       &       &       & [1.121] &       &       & [0.751] & [0.127] & [0.131] \\
   
    Spline2$\times$Ln distance &       &       &       & 3.429*** &       &       & 3.713*** & -0.075 & 0.463*** \\
          &       &       &       & [1.200] &       &       & [0.890] & [0.178] & [0.172] \\
    
    Ln GDP  & 2.122*** & 0.986*** & 0.756*** & 0.806*** & 0.342*** & 0.643*** & 0.677*** & 0.204*** & 0.231*** \\
          & [0.615] & [0.060] & [0.058] & [0.062] & [0.033] & [0.071] & [0.069] & [0.008] & [0.008] \\
    Ln GDP per capita & -0.677 & 0.135 & 0.682*** & 0.694*** & 0.205*** & 0.630*** & 0.600*** & 0.082*** & 0.130*** \\
          & [0.703] & [0.095] & [0.096] & [0.109] & [0.035] & [0.076] & [0.095] & [0.019] & [0.019] \\

    Island &       & -0.342 & -1.047*** & -0.545** & -0.178** & -0.960*** & -0.579*** & -0.007 & 0.003 \\
          &       & [0.343] & [0.305] & [0.241] & [0.080] & [0.212] & [0.175] & [0.041] & [0.041] \\
    Landlocked &       & -0.399 & -0.855*** & -0.824*** & -0.199 & -0.712*** & -0.682*** & 0.143*** & 0.102** \\
          &       & [0.316] & [0.265] & [0.272] & [0.128] & [0.221] & [0.214] & [0.050] & [0.050] \\
    Common religion &       & 0.191 & -0.118 & 0.881* & -0.006 & 0.179 & 0.900*** & 0.335*** & 0.411*** \\
          &       & [0.349] & [0.460] & [0.481] & [0.129] & [0.305] & [0.342] & [0.068] & [0.067] \\
    Common legal origin &       & 0.527* & 0.209 & -0.015 & 0.246** & 0.108 & -0.059 & 0.271*** & 0.184*** \\
          &       & [0.302] & [0.210] & [0.287] & [0.102] & [0.195] & [0.216] & [0.059] & [0.062] \\
    Colony &       & 0.971 & 1.736*** & 1.191*** & 0.521*** & 1.670*** & 1.253*** & 0.140* & 0.174** \\
          &       & [0.691] & [0.303] & [0.328] & [0.185] & [0.235] & [0.222] & [0.073] & [0.078] \\
    
    Constant & -41.120*** & -11.972*** & -13.242*** & 1.126 & -8.602*** & -16.079*** & 5.020 & -8.121*** & -4.970*** \\
          & [9.962] & [2.385] & [1.989] & [11.274] & [1.149] & [2.305] & [6.811] & [1.458] & [1.508] \\

    \midrule
   Fixed Effects   & \multicolumn{1}{c|}{Country} & \multicolumn{3}{c|}{Year}& \multicolumn{3}{c|}{HS8 $\times$ Mode } & \multicolumn{1}{c|}{Year} & HS8  $\times$ Mode\\
   
     & \multicolumn{1}{c|}{} & \multicolumn{3}{c|}{}& \multicolumn{3}{c|}{$\times$ Year } & \multicolumn{1}{c|}{} &  $\times$ Year\\

   \midrule
  SE Clusters   &  \multicolumn{4}{c|}{Country}& \multicolumn{5}{c}{Exporter } \\

    \midrule
    
    Observations & 1,180 & 1,189 & 1,189 & 1,189 & 90,502 & 90,502 & 90,502 & 595,651 & 587,171 \\
    R-squared & 0.966 & 0.768 &       &       & 0.481 &       &       &       &  \\
    Pseudo R-squared & .     & .     & 0.918 & 0.934 & .     & 0.813 & 0.820 & 0.0371 & 0.183 \\
    \bottomrule

    \bottomrule
    \end{tabular}%
  \label{ch2tbase}%
\end{table}%
\vspace{-1em}
\footnotesize{Note: OLS regressions of the estimation equation \ref{ch2cols} in columns 1-2, and \ref{ch2pols} in column 5, where the dependent variable is the log number of shipments. PPML regressions of the estimation equation \ref{ch2cppml} in column 3, \ref{ch2cppmln} in column 4, \ref{ch2pppmln} in columns 6-7, and \ref{ch2fppmln} in columns 8-9, where the dependent variable is the number of shipments. ***$p < 0.01$; **$p < 0.05$; *$p < 0.1$.}
\fillandplacepagenumber
\end{landscape}

\pagebreak
\begin{table}[H]
  \centering
  \tiny
  \singlespacing
  \renewcommand\thetable{6}
  \caption{Shipping Frequency Response for More Strict Fixed Effects}
    \begin{tabular}{lcccccc}
    \toprule
          & (1)   & (2)   & (3)   & (4)   & (5)   & (6) \\
    
    \midrule
         
    Ln per-shipment cost& -0.179*** & -0.192*** & -0.221*** & -0.227*** & -0.266*** & -0.331*** \\
          & [0.025] & [0.028] & [0.028] & [0.030] & [0.028] & [0.031] \\
    Ln distance & -0.479*** & 0.342** & 0.228 & 0.084 & -0.965*** & -1.074*** \\
          & [0.143] & [0.165] & [0.167] & [0.161] & [0.150] & [0.158] \\
    Spline1 & -0.207 & 7.244*** & 5.068*** & 3.467*** & -5.934*** & -6.450*** \\
          & [1.185] & [1.284] & [1.314] & [1.262] & [1.227] & [1.309] \\
    Spline2 & -12.312*** & -0.899 & -2.319 & -3.909** & -17.644*** & -19.400*** \\
          & [1.397] & [1.692] & [1.695] & [1.644] & [1.498] & [1.621] \\
   
    Spline1$\times$Ln distance & -0.094 & -0.836*** & -0.591*** & -0.423*** & 0.517*** & 0.552*** \\
          & [0.125] & [0.133] & [0.136] & [0.131] & [0.129] & [0.138] \\
   
    Spline2$\times$Ln distance & 1.377*** & 0.143 & 0.295* & 0.463*** & 1.939*** & 2.129*** \\
          & [0.146] & [0.178] & [0.178] & [0.172] & [0.157] & [0.170] \\
    Ln GDP  & 0.215*** & 0.223*** & 0.228*** & 0.231*** & 0.247*** & 0.276*** \\
          & [0.007] & [0.008] & [0.008] & [0.008] & [0.008] & [0.009] \\
    Ln GDP per capita & 0.118*** & 0.116*** & 0.126*** & 0.130*** & 0.162*** & 0.185*** \\
          & [0.017] & [0.018] & [0.019] & [0.019] & [0.018] & [0.020] \\
    Island & -0.072** & 0.044 & 0.034 & 0.003 & -0.162*** & -0.221*** \\
          & [0.037] & [0.040] & [0.040] & [0.041] & [0.039] & [0.042] \\
    Landlocked & -0.021 & 0.136*** & 0.116** & 0.102** & -0.087* & -0.116** \\
          & [0.047] & [0.050] & [0.050] & [0.050] & [0.050] & [0.053] \\
    Common religion & 0.443*** & 0.389*** & 0.413*** & 0.411*** & 0.490*** & 0.508*** \\
          & [0.055] & [0.067] & [0.067] & [0.067] & [0.060] & [0.065] \\
    Common legal origin & 0.213*** & 0.214*** & 0.189*** & 0.184*** & 0.126** & 0.105* \\
          & [0.050] & [0.061] & [0.062] & [0.062] & [0.053] & [0.056] \\
    Colony & 0.055 & 0.091 & 0.138* & 0.174** & 0.252*** & 0.322*** \\
          & [0.068] & [0.076] & [0.078] & [0.078] & [0.072] & [0.075] \\
    Constant & 0.690 & -7.407*** & -6.291*** & -4.970*** & 4.863*** & 5.543*** \\
          & [1.308] & [1.535] & [1.560] & [1.508] & [1.368] & [1.446] \\

    \midrule
      Fixed Effects & Exporter   & HS8   & HS8 $\times$ Mode   & HS8 $\times$ Mode   & HS8 $\times$ Mode   & HS4 $\times$ Mode \\
        &   &    &   & $\times$ Year   & $\times$ Year + Exporter    & $\times$ Year $\times$ Exporter \\
    \midrule
    Observations & 591,808 & 594,533 & 593,352 & 587,171 & 583,315 & 480,657 \\
    Pseudo R-squared & 0.254 & 0.143 & 0.169 & 0.183 & 0.384 & 0.492 \\
    \bottomrule
    \bottomrule
    \end{tabular}%
  \label{ch2fe}%
\end{table}%
\vspace{-1em}
\footnotesize{Note: PPML regressions of the estimation equation \ref{ch2fppmln}. The dependent variable is the number of shipments. Standard errors are robust and clustered with exporter clusters. ***$p < 0.01$; **$p < 0.05$; *$p < 0.1$.}

\pagebreak
\begin{landscape}

\begin{table}[H]
  \centering
  \tiny
  \singlespacing
  \renewcommand\thetable{7}
  \caption{Financing Costs and Shipping Frequency}
    \begin{tabular}{lccccccc}
    \toprule
    &    &   & & & Air   &  Ocean  & Land   \\
    \cmidrule(lr){6-8}
          & (1)   & (2)   & (3)   & (4)   & (5)   & (6) & (7) \\
    
    \midrule
          
    Ln per-shipment cost& -0.196*** & -0.159*** & -0.198*** & -0.335*** & -0.310*** & -0.352*** & -0.888*** \\
          & [0.034] & [0.032] & [0.034] & [0.038] & [0.072] & [0.038] & [0.285] \\
    Ln distance & 0.569*** & 0.992*** & 0.551*** & -0.904*** & -0.319 & -0.973*** & -73.982*** \\
          & [0.179] & [0.194] & [0.178] & [0.179] & [0.340] & [0.191] & [20.103] \\
    Spline1 & 5.652*** & 9.487*** & 5.488*** & -5.023*** & 1.781 & -105.202*** &  \\
          & [1.386] & [1.493] & [1.381] & [1.423] & [3.383] & [31.736] &  \\
    Spline2 & 1.509 & 5.275** & 1.322 & -19.402*** & -13.294*** & -19.924*** &  \\
          & [1.977] & [2.108] & [1.964] & [1.977] & [3.940] & [2.087] &  \\
   
    Spline1$\times$Ln distance & -0.584*** & -0.980*** & -0.568*** & 0.410*** & -0.298 & 0.510*** &  \\
          & [0.145] & [0.154] & [0.144] & [0.149] & [0.368] & [0.151] &  \\
   
    Spline2$\times$Ln distance & -0.107 & -0.500** & -0.087 & 2.135*** & 1.484*** & 2.190*** &  \\
          & [0.208] & [0.221] & [0.206] & [0.208] & [0.421] & [0.219] &  \\
   Importer interest rate   & -0.004** & -0.016*** & -0.043*** & -0.083*** & 0.006 & -0.104*** & 1.877*** \\
          & [0.002] & [0.002] & [0.014] & [0.018] & [0.027] & [0.019] & [0.348] \\
    Exporter interest rate &       & 0.036*** &       &       &       &       &  2.094***\\
          &       & [0.007] &       &       &       &       &  [0.322]\\
    Importer interest rate $\times$Exporter interest rate&       &       & 0.003*** & 0.006*** & -0.001 & 0.007*** &  -0.165***\\
          &       &       & [0.001] & [0.001] & [0.002] & [0.001] &  [0.029]\\
          
    Ln GDP  & 0.230*** & 0.222*** & 0.230*** & 0.300*** & 0.294*** & 0.301*** & -14.041*** \\
          & [0.008] & [0.008] & [0.008] & [0.010] & [0.024] & [0.010] & [3.283] \\
    Ln GDP per capita & 0.157*** & 0.130*** & 0.157*** & 0.248*** & 0.282*** & 0.232*** & 14.426*** \\
          & [0.020] & [0.020] & [0.020] & [0.025] & [0.048] & [0.025] & [3.424] \\
    Island & -0.050 & -0.032 & -0.051 & -0.234*** & -0.201** & -0.240*** &  \\
          & [0.041] & [0.041] & [0.041] & [0.046] & [0.078] & [0.051] &  \\
    Landlocked & 0.081 & 0.087* & 0.082 & -0.162*** & -0.234 & -0.147*** &  \\
          & [0.050] & [0.050] & [0.050] & [0.057] & [0.184] & [0.055] &  \\
    Common religion & 0.488*** & 0.514*** & 0.490*** & 0.789*** & 1.025*** & 0.663*** &  \\
          & [0.103] & [0.102] & [0.103] & [0.117] & [0.239] & [0.115] &  \\
    Common legal origin & 0.269*** & 0.288*** & 0.268*** & 0.081 & -0.053 & 0.117 &  \\
          & [0.071] & [0.072] & [0.071] & [0.069] & [0.092] & [0.075] &  \\
    Colony & 0.167** & 0.082 & 0.169** & 0.359*** & 0.500*** & 0.326*** &  \\
          & [0.077] & [0.078] & [0.077] & [0.082] & [0.107] & [0.090] &  \\

    Constant & -10.105*** & -14.318*** & -9.914*** & 2.859* & -3.450 & 3.830** & 918.859*** \\
          & [1.716] & [1.885] & [1.710] & [1.666] & [3.032] & [1.765] & [244.615] \\

    \midrule
      Fixed Effects & HS8 $\times$ Mode  & HS8  $\times$    & HS8 $\times$ Mode  & Exporter $\times$ HS8    & Exporter $\times$    & Exporter $\times$  &Exporter  \\
        &  $\times$ Year  & Mode  & Year  & Mode$\times$ Year   &  HS8  $\times$ Year  & HS8  $\times$ Year & $\times$ Year\\
    \midrule
    
    Observations & 539,257 & 545,159 & 539,257 & 371,871 & 92,077 & 279,790 & 4,643 \\
    Pseudo R-squared & 0.184 & 0.172 & 0.184 & 0.564 & 0.523 & 0.567 & 0.681 \\
    \bottomrule

    \bottomrule
    \end{tabular}%
  \label{ch2fin}%
\end{table}%
\vspace{-1em}
\footnotesize{Note: PPML regressions of the estimation equation \ref{ch2fppmlnf}. The dependent variable is the number of shipments. Standard errors are robust and clustered with exporter clusters. ***$p < 0.01$; **$p < 0.05$; *$p < 0.1$.}
\fillandplacepagenumber
\end{landscape}

\pagebreak
\begin{table}[H]
  \centering
  \tiny
  \singlespacing
  \renewcommand\thetable{8}
  \caption{Alternative Financing Costs}
    \begin{tabular}{lcccc}
    \toprule
          & (1)   & (2)   & (3)   & (4) \\
    
    \midrule

   Ln per-shipment cost& -0.250*** & -0.430*** & -0.347*** & -0.389*** \\
          & [0.035] & [0.050] & [0.042] & [0.035] \\
    Ln distance & -1.017*** & -1.635*** & -2.257*** & -1.636*** \\
          & [0.186] & [0.255] & [0.291] & [0.244] \\
    Spline1 & -5.804*** & -9.817*** & -12.037*** & -9.541*** \\
          & [1.429] & [1.775] & [1.946] & [1.702] \\
    Spline2 & -21.049*** & -17.129*** & -30.206*** & -21.007*** \\
          & [2.073] & [2.294] & [3.118] & [2.496] \\
    
    Spline1$\times$Ln distance & 0.494*** & 0.896*** & 1.037*** & 0.828*** \\
          & [0.149] & [0.182] & [0.196] & [0.175] \\
    
    Spline2$\times$Ln distance & 2.311*** & 1.812*** & 3.227*** & 2.244*** \\
          & [0.218] & [0.240] & [0.325] & [0.261] \\
   Importer interest rate   & -0.078*** & -0.142*** & -0.130*** &  \\
          & [0.018] & [0.021] & [0.020] &  \\

    Importer interest rate $\times$Exporter interest rate& 0.006*** & 0.010*** & 0.009*** &  \\
          & [0.001] & [0.002] & [0.002] &  \\
    Access to credit & 0.026*** &       &       &  \\
          & [0.007] &       &       &  \\
    
    Bank confidence &       & 0.444*** &       &  \\
          &       & [0.092] &       &  \\
    Debt-GDP ratio     &       &       & -0.003*** &  \\
          &       &       & [0.000] &  \\
    Importer net interest margin   &       &       &       & 0.069*** \\
          &       &       &       & [0.025] \\
   
    Importer net interest margin$\times$Exporter net interest margin &       &       &       & -0.021*** \\
          &       &       &       & [0.006] \\
          
    Ln GDP  & 0.290*** & 0.283*** & 0.311*** & 0.281*** \\
          & [0.010] & [0.012] & [0.013] & [0.011] \\
    Ln GDP per capita & 0.248*** & 0.319*** & 0.529*** & 0.399*** \\
          & [0.025] & [0.035] & [0.036] & [0.030] \\
    Island & -0.226*** & -0.536*** & -0.527*** & -0.664*** \\
          & [0.046] & [0.044] & [0.044] & [0.048] \\
    Landlocked & -0.231*** & 1.295*** & 1.462*** & 0.571*** \\
          & [0.062] & [0.368] & [0.363] & [0.214] \\
    Common religion & 0.873*** & 0.486*** & 0.669*** & 0.842*** \\
          & [0.119] & [0.125] & [0.140] & [0.086] \\
    Common legal origin & -0.038 & 0.024 & -0.133 & -0.085 \\
          & [0.080] & [0.089] & [0.105] & [0.088] \\
    Colony & 0.322*** & 0.581*** & 0.670*** & 0.556*** \\
          & [0.082] & [0.170] & [0.166] & [0.161] \\
    
    Constant & 3.326** & 9.430*** & 13.439*** & 9.579*** \\
          & [1.688] & [2.351] & [2.701] & [2.237] \\
     \midrule
    Observations & 371,871 & 175,814 & 173,652 & 196,751 \\
    Pseudo R-squared & 0.564 & 0.635 & 0.634 & 0.619 \\
    \bottomrule

    \bottomrule
    \end{tabular}%
  \label{ch2finalt}%
\end{table}%
\vspace{-1em}
\footnotesize{Note: PPML regressions of the estimation equation \ref{ch2fppmlnf}. The dependent variable is the number of shipments. Exporter by product by mode by year fixed effects is used in all regressions. Standard errors are robust and clustered with exporter clusters. ***$p < 0.01$; **$p < 0.05$; *$p < 0.1$.}

\pagebreak
\begin{table}[H]
  \centering
  \tiny
  \singlespacing
  \renewcommand\thetable{9}
  \caption{Robustness Check: Alternative Distance and Time}
    \begin{tabular}{lcccc}
    \toprule
          & (1)   & (2)   & (3)   & (4) \\
    
    \midrule

    Ln per-shipment cost& -0.233*** & -0.163*** & -0.352*** & -0.273*** \\
          & [0.039] & [0.040] & [0.039] & [0.045] \\
    Ln distance & -1.195*** & -1.642*** & -0.976*** & -0.655*** \\
          & [0.203] & [0.225] & [0.193] & [0.209] \\
    Spline1 & -6.147*** & -11.686*** & -5.878*** & -4.311*** \\
          & [1.573] & [1.748] & [1.477] & [1.539] \\
    Spline2 & -21.370*** & -24.908*** & -20.008*** & -15.835*** \\
          & [2.224] & [2.394] & [2.144] & [2.360] \\
    
    Spline1$\times$Ln distance & 0.574*** & 1.204*** & 0.497*** & 0.362** \\
          & [0.164] & [0.185] & [0.153] & [0.158] \\
    
    Spline2$\times$Ln distance & 2.369*** & 2.752*** & 2.199*** & 1.754*** \\
          & [0.234] & [0.253] & [0.226] & [0.248] \\
    Ln sea-distance & 0.553*** &       &       &  \\
          & [0.078] &       &       &  \\
          
    Transport time (days)  &       & 0.035*** &       &  \\
          &       & [0.004] &       &  \\
    Total transport time (days) &       &       & 0.000 &  \\
          &       &       & [0.003] &  \\
    Logistic Performance Index   &       &       &       & 0.259*** \\
          &       &       &       & [0.072] \\
          
   Importer interest rate   & -0.092*** & -0.085*** & -0.103*** & -0.097*** \\
          & [0.019] & [0.019] & [0.019] & [0.018] \\
   
    Importer interest rate $\times$Exporter interest rate& 0.007*** & 0.006*** & 0.007*** & 0.007*** \\
          & [0.001] & [0.001] & [0.001] & [0.001] \\
    Ln GDP  & 0.312*** & 0.319*** & 0.302*** & 0.290*** \\
          & [0.010] & [0.011] & [0.010] & [0.011] \\
    Ln GDP per capita & 0.219*** & 0.201*** & 0.237*** & 0.149*** \\
          & [0.025] & [0.025] & [0.027] & [0.036] \\
    Island & -0.034 & -0.050 & -0.237*** & -0.270*** \\
          & [0.062] & [0.056] & [0.060] & [0.052] \\
    Landlocked & -0.038 & -0.007 & -0.150*** & -0.145*** \\
          & [0.056] & [0.057] & [0.057] & [0.056] \\
    Common religion & 1.220*** & 0.994*** & 0.680*** & 0.749*** \\
          & [0.133] & [0.119] & [0.121] & [0.119] \\
    Common legal origin & 0.225*** & 0.244*** & 0.112 & 0.168** \\
          & [0.075] & [0.074] & [0.077] & [0.075] \\
    Colony & -0.039 & -0.062 & 0.327*** & 0.296*** \\
          & [0.106] & [0.098] & [0.099] & [0.091] \\
    
    Constant & -0.617 & 7.445*** & 3.771** & 0.429 \\
          & [1.909] & [1.982] & [1.785] & [1.957] \\

    \midrule
     Fixed Effects &   Exporter$\times$ HS8     &  Exporter$\times$ HS8     & Exporter$\times$ HS8      & Exporter $\times$ HS8\\
      &   $\times$ Year    &   $\times$ Year    &   $\times$ Year    & Mode $\times$ Year \\
    \midrule
    Observations & 278,922 & 278,928 & 278,928 & 278,928 \\
    Pseudo R-squared & 0.569 & 0.569 & 0.567 & 0.567 \\
   
    \bottomrule
    \bottomrule
    \end{tabular}%
  \label{ch2time}%
\end{table}%
\vspace{-1em}
\footnotesize{Note: PPML regressions of the estimation equation \ref{ch2fppmlnf}. The dependent variable is the number of shipments. Standard errors are robust and clustered with exporter clusters. ***$p < 0.01$; **$p < 0.05$; *$p < 0.1$.}

\pagebreak
\begin{landscape}

\begin{table}[H]
  \centering
  \tiny
  \singlespacing
  \renewcommand\thetable{10}
  \caption{Robustness Check: Standard Errors Clustering}
    \begin{tabular}{lcccccc}
    \toprule
          & (1)   & (2)   & (3)   & (4)   & (5)   & (6) \\
    
    \midrule
         
   Ln per-shipment cost& -0.335*** & -0.335*** & -0.335*** & -0.335*** & -0.335*** & -0.335*** \\
          & [0.119] & [0.058] & [0.058] & [0.052] & [0.036] & [0.020] \\
    Ln distance & -0.904** & -0.904*** & -0.904*** & -0.904*** & -0.904*** & -0.904*** \\
          & [0.415] & [0.152] & [0.313] & [0.244] & [0.173] & [0.107] \\
    Spline1 & -5.023 & -5.023*** & -5.023* & -5.023** & -5.023*** & -5.023*** \\
          & [3.929] & [1.949] & [2.935] & [2.318] & [1.400] & [0.895] \\
    Spline2 & -19.402*** & -19.402*** & -19.402*** & -19.402*** & -19.402*** & -19.402*** \\
          & [6.117] & [2.375] & [3.560] & [2.764] & [1.907] & [1.170] \\
    
    Spline1$\times$Ln distance & 0.410 & 0.410* & 0.410 & 0.410 & 0.410*** & 0.410*** \\
          & [0.433] & [0.218] & [0.319] & [0.252] & [0.147] & [0.094] \\
    
    Spline2$\times$Ln distance & 2.135*** & 2.135*** & 2.135*** & 2.135*** & 2.135*** & 2.135*** \\
          & [0.658] & [0.265] & [0.380] & [0.295] & [0.201] & [0.123] \\
   Importer interest rate   & -0.083* & -0.083*** & -0.083* & -0.083** & -0.083*** & -0.083*** \\
          & [0.044] & [0.024] & [0.044] & [0.039] & [0.015] & [0.018] \\
   
    Importer interest rate $\times$Exporter interest rate& 0.006* & 0.006*** & 0.006* & 0.006** & 0.006*** & 0.006*** \\
          & [0.003] & [0.002] & [0.003] & [0.003] & [0.001] & [0.001] \\
    Ln GDP  & 0.300*** & 0.300*** & 0.300*** & 0.300*** & 0.300*** & 0.300*** \\
          & [0.032] & [0.029] & [0.014] & [0.018] & [0.009] & [0.005] \\
    Ln GDP per capita & 0.248*** & 0.248*** & 0.248*** & 0.248*** & 0.248*** & 0.248*** \\
          & [0.073] & [0.031] & [0.033] & [0.026] & [0.020] & [0.011] \\
    Island & -0.234* & -0.234*** & -0.234*** & -0.234*** & -0.234*** & -0.234*** \\
          & [0.123] & [0.046] & [0.055] & [0.045] & [0.045] & [0.024] \\
    Landlocked & -0.162 & -0.162** & -0.162 & -0.162** & -0.162*** & -0.162*** \\
          & [0.245] & [0.077] & [0.102] & [0.077] & [0.053] & [0.029] \\
    Common religion & 0.789** & 0.789*** & 0.789*** & 0.789*** & 0.789*** & 0.789*** \\
          & [0.314] & [0.183] & [0.152] & [0.129] & [0.092] & [0.054] \\
    Common legal origin & 0.081 & 0.081 & 0.081 & 0.081 & 0.081 & 0.081** \\
          & [0.157] & [0.056] & [0.078] & [0.060] & [0.058] & [0.033] \\
    Colony & 0.359** & 0.359*** & 0.359*** & 0.359*** & 0.359*** & 0.359*** \\
          & [0.168] & [0.083] & [0.077] & [0.060] & [0.075] & [0.040] \\
    Constant & 2.859 & 2.859 & 2.859 & 2.859 & 2.859* & 2.859*** \\
          & [4.017] & [2.075] & [2.902] & [2.279] & [1.680] & [1.045] \\

    \midrule
    SE Clusters &    Country   &  Exporter     &  Country     &   Country $\times$   &    Country $\times$  & Exporter $\times$ HS2 \\
      &       &      &    $\times$ Year     &    HS2 $\times$ Year    &   HS2 $\times$ Exporter     &  $\times$ Country $\times$ Year\\
    \midrule
         
   Observations & 371,871 & 371,871 & 371,871 & 371,871 & 371,871 & 371,871 \\
    Pseudo R-squared & 0.564 & 0.564 & 0.564 & 0.564 & 0.564 & 0.564 \\
    \bottomrule
    \bottomrule
    \end{tabular}%
  \label{ch2se}%
\end{table}%
\vspace{-1em}
\footnotesize{Note: PPML regressions of the estimation equation \ref{ch2fppmlnf}. The dependent variable is the number of shipments. Exporter by product by mode by year fixed effects is used in all regressions. ***$p < 0.01$; **$p < 0.05$; *$p < 0.1$.}
\fillandplacepagenumber
\end{landscape}

\pagebreak
\begin{table}[H]
  \centering
  \tiny
  \singlespacing
  \renewcommand\thetable{11}
  \caption{Robustness Check: Subsamples}
    \begin{tabular}{lccccc}
    \toprule
          & (1)   & (2)   & (3)   & (4)   & (5) \\
    
    \midrule
         
    Ln per-shipment cost& -0.344*** & -0.399*** & -0.343*** & -0.159*** & -0.336*** \\
          & [0.038] & [0.039] & [0.039] & [0.021] & [0.038] \\
    Ln distance & -0.792*** & -0.982*** & -0.733*** & -0.971*** & -0.856*** \\
          & [0.190] & [0.189] & [0.189] & [0.096] & [0.178] \\
    Spline1 & -4.376*** & -6.618*** & -4.644*** & -4.990*** & -4.365*** \\
          & [1.523] & [1.445] & [1.512] & [0.920] & [1.438] \\
    Spline2 & -17.968*** & -19.697*** & -17.014*** & -16.062*** & -18.926*** \\
          & [2.090] & [2.075] & [2.091] & [1.153] & [1.972] \\
    
    Spline1$\times$Ln distance & 0.355** & 0.589*** & 0.395** & 0.424*** & 0.340** \\
          & [0.160] & [0.150] & [0.159] & [0.099] & [0.152] \\
    
    Spline2$\times$Ln distance & 1.984*** & 2.165*** & 1.883*** & 1.742*** & 2.083*** \\
          & [0.220] & [0.218] & [0.220] & [0.123] & [0.207] \\
   Importer interest rate   & -0.079*** & -0.126*** & -0.076*** & -0.054*** & -0.083*** \\
          & [0.018] & [0.024] & [0.019] & [0.011] & [0.018] \\
    
    Importer interest rate $\times$Exporter interest rate& 0.006*** & 0.010*** & 0.006*** & 0.004*** & 0.006*** \\
          & [0.001] & [0.002] & [0.001] & [0.001] & [0.001] \\
    Ln GDP  & 0.294*** & 0.300*** & 0.278*** & 0.197*** & 0.304*** \\
          & [0.010] & [0.010] & [0.011] & [0.005] & [0.010] \\
    Ln GDP per capita & 0.231*** & 0.239*** & 0.220*** & 0.208*** & 0.244*** \\
          & [0.025] & [0.025] & [0.026] & [0.014] & [0.025] \\
    Island & -0.225*** & -0.240*** & -0.252*** & -0.113*** & -0.218*** \\
          & [0.047] & [0.043] & [0.047] & [0.024] & [0.045] \\
    Landlocked & -0.142** & -0.127*** & -0.122** & -0.174*** & -0.173*** \\
          & [0.057] & [0.048] & [0.059] & [0.034] & [0.057] \\
    Common religion & 0.728*** & 0.685*** & 0.778*** & 0.607*** & 0.807*** \\
          & [0.119] & [0.127] & [0.124] & [0.073] & [0.114] \\
    Common legal origin & 0.107 & 0.128* & 0.135* & -0.108*** & 0.077 \\
          & [0.071] & [0.076] & [0.072] & [0.028] & [0.069] \\
    Colony & 0.325*** & 0.345*** & 0.303*** & 0.318*** & 0.338*** \\
          & [0.083] & [0.086] & [0.084] & [0.037] & [0.082] \\
    Constant & 2.215 & 4.194** & 2.399 & 5.068*** & 2.364 \\
          & [1.760] & [1.771] & [1.770] & [0.934] & [1.661] \\
     \midrule
    Observations & 333,132 & 228,524 & 218,642 & 366,099 & 375,216 \\
    Pseudo R-squared & 0.565 & 0.581 & 0.552 & 0.447 & 0.565 \\
    \bottomrule
    
    \bottomrule
    \end{tabular}%
  \label{ch2sub}%
\end{table}%
\vspace{-1em}
\footnotesize{Note: PPML regressions of the estimation equation \ref{ch2fppmlnf}. The dependent variable is the number of shipments. Exporter by product by mode fixed effects is used in all regressions. Standard errors are robust and clustered with exporter clusters. ***$p < 0.01$; **$p < 0.05$; *$p < 0.1$.}

\pagebreak

\section{Appendix}
\subsection{Appendix-A} \label{a1}
\subsection*{Propositions}

\hspace{.5cm}\textbf{Proof of Proposition 1:}

Rewriting the objective function
\begin{equation*}
   \min_{x} \dfrac{(cx\mathrm{e}^{ \Delta r}+f)\mathrm{e}^{- \Delta r}}{1-\mathrm{e}^{-\frac{r_1x}{q}}}
\end{equation*}

Taking the derivative of the objective function for $x$, the first-order condition (FOC) is
\begin{equation*}
    \dfrac{c\mathrm{e}^{{\Delta}r-{\Delta}r_1}}{1-\mathrm{e}^{-\frac{r_1x}{q}}}-\dfrac{r_1\left(c\mathrm{e}^{{\Delta}r}x+f\right)\mathrm{e}^{-\frac{r_1x}{q}-{\Delta}r_1}}{q\left(1-\mathrm{e}^{-\frac{r_1x}{q}}\right)^2}
\end{equation*}

For cost minimization, I look at the second-order condition (SOC)
\begin{equation*}
    \begin{multlined}
    -\dfrac{cr_1\mathrm{e}^{-\frac{r_1x}{q}-{\Delta}r_1+{\Delta}r}}{q\left(1-\mathrm{e}^{-\frac{r_1x}{q}}\right)^2}+\dfrac{r_1^2\left(c\mathrm{e}^{{\Delta}r}x+f\right)\mathrm{e}^{-\frac{r_1x}{q}-{\Delta}r_1}}{q^2\left(1-\mathrm{e}^{-\frac{r_1x}{q}}\right)^2}+\dfrac{2r_1^2\left(c\mathrm{e}^{{\Delta}r}x+f\right)\mathrm{e}^{-\frac{2r_1x}{q}-{\Delta}r_1}}{q^2\left(1-\mathrm{e}^{-\frac{r_1x}{q}}\right)^3}-\dfrac{cr_1\mathrm{e}^{{\Delta}r-{\Delta}r_1}\mathrm{e}^{-\frac{r_1x}{q}}}{q\left(1-\mathrm{e}^{-\frac{r_1x}{q}}\right)^2}
    \end{multlined}
\end{equation*}

I simplify SOC by rearranging the above condition 
\begin{equation*}
   SOC = \dfrac{r_1 e^{\frac{r_1x}{q}-\Delta r_1}}{q^2(e^\frac{r_1x}{q}-1)^3}\bigg((ce^{\Delta r}r_1x+fr_1)(e^\frac{r_1x}{q}+1)-2cqe^{\Delta r}(e^{\frac{r_1 x}{q}}-1)\bigg)
\end{equation*}

By using equation \ref{eq:3}, I get 
\begin{equation*}
    SOC= \frac{r_1 e^{\frac{r_1x}{q}-\Delta r_1}}{q^2(e^\frac{r_1x}{q}-1)^3}\bigg((ce^{\Delta r}r_1x+fr_1)(e^\frac{r_1x}{q}-1)\bigg)
\end{equation*}

By using equation \ref{eq:3} again, 
\begin{equation*}
    SOC = \frac{c r_1 e^{\frac{r_1x}{q}-\Delta r_1+\Delta r}}{q(e^\frac{r_1x}{q}-1)}
\end{equation*}

For $x>0$,
\begin{equation*}
    SOC >0
\end{equation*}

The SOC is strictly positive, so I have a unique minimizer $x^*$. To strengthen this case, let's consider the FOC again,
\begin{equation*} 
    \big[qc \mathrm{e}^{\Delta r} (\mathrm{e}^{\frac{r_1 x}{q}}-1) - c \mathrm{e}^{\Delta r} r_1 x \big]- f r_1
\end{equation*}

At the limit, if $x=0$, the square bracketed term is zero, and the above statement is negative. Taking the derivative of the square bracketed term concerning $x$ gives
\begin{equation*} 
    c r_1 \mathrm{e}^{\Delta r} (\mathrm{e}^{\frac{r_1 x}{q}}) - c \mathrm{e}^{\Delta r} r_1 
\end{equation*}

Rearranging,
\begin{equation*} 
    c r_1 \mathrm{e}^{\Delta r} (\mathrm{e}^{\frac{r_1 x}{q}}-1) \geq 0
\end{equation*}

This shows that the square bracketed term is increasing in $x$. When $x$ approaches $q$, the statement is positive. Thus, there must be a unique $x^*$ where FOC is zero.

\vspace{1cm}
\textbf{Proof of Proposition 2:}

\textbf{(i)} Taking the derivative with respect to $r$
\begin{equation*}
    \frac{\partial FOC}{\partial r}=\dfrac{c{\Delta}\left(q\mathrm{e}^\frac{r_1x}{q}-r_1x-q\right)\mathrm{e}^{{\Delta}r+\frac{r_1x}{q}-{\Delta}r_1}}{q\left(\mathrm{e}^\frac{r_1x}{q}-1\right)^2}
\end{equation*}

Here, I use inequality of $\mathrm{e}^{x}\geq 1+x$. In this case, I have $\mathrm{e}^{\frac{r_1x}{q}}-1 \geq \frac{r_1x}{q}$. Rearranging, $q(\mathrm{e}^{\frac{r_1x}{q}}-1) -r_1x \geq 0$. Using the above statement, I get $ \frac{\partial FOC}{\partial r}\geq 0$. 

Implicit function theorem for r gives, $\frac{\partial x}{\partial r} = \frac{- \frac{\partial FOC}{\partial r}}{SOC}\leq 0$. The exporter chooses a small shipment size for high financing costs in the exporter country. 

For the relationship between n and r, I use $n=\dfrac{q}{x}$. Taking derivative with respect to $r$ and using $\frac{\partial x}{\partial r} \leq 0$ give
\begin{equation*}
    \dfrac{\partial n}{\partial r} = - \dfrac{q}{x(.)^2} \dfrac{\partial x(.)}{\partial r}\geq 0
\end{equation*}

\vspace{1cm}
\textbf{(ii)} Taking derivative of FOC with respect to $r_1$
\begin{align*}
    \dfrac{\partial FOC}{\partial r_1}= -\dfrac{cx\mathrm{e}^{-\frac{xr_1}{q}-{\Delta}r_1+{\Delta}r}}{q\left(1-\mathrm{e}^{-\frac{xr_1}{q}}\right)^2}-\dfrac{\left(-\frac{x}{q}-{\Delta}\right)\left(c\mathrm{e}^{{\Delta}r}x+f\right)r_1\mathrm{e}^{-\frac{xr_1}{q}-{\Delta}r_1}}{q\left(1-\mathrm{e}^{-\frac{xr_1}{q}}\right)^2}-\dfrac{\left(c\mathrm{e}^{{\Delta}r}x+f\right)\mathrm{e}^{-\frac{xr_1}{q}-{\Delta}r_1}}{q\left(1-\mathrm{e}^{-\frac{xr_1}{q}}\right)^2}\\+\dfrac{2x\left(c\mathrm{e}^{{\Delta}r}x+f\right)r_1\mathrm{e}^{-\frac{2xr_1}{q}-{\Delta}r_1}}{q^2\left(1-\mathrm{e}^{-\frac{xr_1}{q}}\right)^3}-\dfrac{c{\Delta}\mathrm{e}^{{\Delta}r-{\Delta}r_1}}{1-\mathrm{e}^{-\frac{xr_1}{q}}}
\end{align*}
 
By using equation \ref{eq:3}, 
\begin{equation*}
    \dfrac{\partial FOC}{\partial r_1}=\dfrac{c \mathrm{e}^{\Delta r-\Delta r_1}}{1-\mathrm{e}^{\frac{-r_1 x}{q}}} \bigg[\frac{-x \mathrm{e}^{\frac{-r_1 x}{q}}}{q (1-\mathrm{e}^{\frac{-r_1 x}{q}})}+\dfrac{x}{q}+\Delta-\dfrac{1}{r_1}-\Delta+\dfrac{2x }{q (1-\mathrm{e}^{\frac{-r_1 x}{q}})} \bigg]
\end{equation*}
  
After manipulation

\begin{equation*}
    \dfrac{\partial FOC}{\partial r_1}=\dfrac{c \mathrm{e}^{\Delta r-\Delta r_1}}{1-\mathrm{e}^{\frac{-r_1 x}{q}}} \bigg[\frac{x (3-2\mathrm{e}^{\frac{-r_1 x}{q}})}{q (1-\mathrm{e}^{\frac{-r_1 x}{q}})}-\dfrac{1}{r_1}\bigg]
\end{equation*}

By using inequality property of $1-e^{\frac{-r_1x}{q}} \leq \frac{r_1x}{q}$, I get $\frac{x}{q (1-\mathrm{e}^{\frac{-r_1 x}{q}})} \geq \frac{1}{r_1}$. Using $(3-2\mathrm{e}^{\frac{-r_1 x}{q}})>1$, the square bracketed term is positive. From the equation above, I get $\frac{\partial FOC}{\partial r_1}\geq 0$.

Implicit function theorem for $r_1$ gives $\frac{\partial x}{\partial r_1} = \frac{- \frac{\partial FOC}{\partial r_1}}{SOC}$. By using $SOC>0$ and $\frac{\partial FOC}{\partial r_1}\geq 0$, I get $\frac{\partial x}{\partial r_1}\leq 0.$ The importer chooses a small shipment size for high financing costs in the importer country. 

Using the relationship between n and r, and using $\frac{\partial x}{\partial r} \leq 0$ give
\begin{equation*}
    \dfrac{\partial n}{\partial r_1} = - \dfrac{q}{x(.)^2} \dfrac{\partial x(.)}{\partial r_1}\geq 0
\end{equation*}

\textbf{(iii)} Taking the derivative of FOC with respect to $\Delta$
\begin{equation*}
     \dfrac{\partial FOC}{\partial \Delta} = -\dfrac{\left(\left(\left(cqr_1-cqr\right)\mathrm{e}^\frac{r_1x}{q}+\left(crr_1-cr_1^2\right)x-cqr_1+cqr\right)\mathrm{e}^{r{\Delta}}-fr_1^2\right)\mathrm{e}^{\frac{r_1x}{q}-{\Delta} r_1}}{q\left(\mathrm{e}^\frac{r_1x}{q}-1\right)^2}
\end{equation*}

Rearranging and by using equation \ref{eq:3},
\begin{equation*}
     \dfrac{\partial FOC}{\partial \Delta} =\dfrac{frr_1\mathrm{e}^{\frac{r_1x}{q}-{\Delta}r_1}}{q\left(\mathrm{e}^\frac{r_1x}{q}-1\right)^2} \geq 0
\end{equation*}

Implicit function theorem for $\Delta$ gives $\frac{\partial x}{\partial \Delta} = \frac{- \frac{\partial FOC}{\partial \Delta}}{SOC} \leq 0$. 

Using the relationship between n and $\Delta$ and using $\frac{\partial x}{\partial \Delta}\leq 0$, I have 
\begin{equation*}
    \frac{\partial n}{\partial \Delta} = - \frac{q}{x(.)^2} \frac{\partial x(.)}{\partial \Delta}\geq 0
\end{equation*}

\textbf{(iv)} Taking derivative of FOC with respect to $f$
\begin{equation*}
    \dfrac{\partial FOC}{\partial f} = -\dfrac{r_1\mathrm{e}^{\frac{r_1x}{q}-{\Delta}r_1}}{q\left(\mathrm{e}^\frac{r_1x}{q}-1\right)^2} \leq 0
\end{equation*}

Implicit function theorem for $f$ gives, $\frac{\partial x}{\partial f} = \frac{- \frac{\partial FOC}{\partial f}}{SOC}$. By using $\dfrac{\partial FOC}{\partial f} \leq 0$, I get
\begin{equation} \label{eqpc1} 
    \frac{\partial x}{\partial f} \geq 0
\end{equation}

For the relationship between n and $f$, I have 
\begin{equation*}
    \dfrac{\partial n}{\partial f} = - \dfrac{q}{x(.)^2} \dfrac{\partial x(.)}{\partial f}
\end{equation*}

By using $\frac{\partial x}{\partial f} \geq 0$ from equation \ref{eqpc1} gives
\begin{equation*}
    \dfrac{\partial n}{\partial f} \leq 0
\end{equation*}

\vspace{1cm}
\textbf{Proof of Proposition 3:}

\textbf{a)}

\textbf{(i-iii)}
The exporter's financing demand is $D=cx^*(.)$. I take derivative of $D$ with respect to $r$, $r_1$, and $\Delta$ 

\begin{equation*}
    \dfrac{\partial D}{\partial r}=  c \frac{\partial x}{\partial r} 
\end{equation*}
\begin{equation*}
    \dfrac{\partial D}{\partial r_1}=  c \frac{\partial x}{\partial r_1}
\end{equation*}
\begin{equation*}
    \dfrac{\partial D}{\partial \Delta}= c \frac{\partial x}{\partial \Delta}
\end{equation*}

By using $\frac{\partial x}{\partial r} \leq 0$, $\frac{\partial x}{\partial r_1} \leq 0$, and $\frac{\partial x}{\partial \Delta} \leq 0$, I find

\begin{equation*}
    \dfrac{\partial D}{\partial r} \leq 0
\end{equation*}
\begin{equation*}
    \dfrac{\partial D}{\partial r_1} \leq 0
\end{equation*}
\begin{equation*}
    \dfrac{\partial D}{\partial \Delta} \leq 0
\end{equation*}

\textbf{(iv)}
I take derivative of $D$ with respect to $f$
\begin{equation*}
    \dfrac{\partial D}{\partial f}= c \frac{\partial x}{\partial f} 
\end{equation*}

By using $\frac{\partial x}{\partial f} \geq 0$ from equation \ref{eqpc1}, I find
\begin{equation*}
    \dfrac{\partial D}{\partial f} \geq 0
\end{equation*}

\textbf{b)}

For $r$, 
\begin{equation*}
    \dfrac{\partial^2 D}{\partial r \partial f}=  c \frac{\partial^2 x}{\partial r \partial f} 
\end{equation*}

We need to sign $\frac{\partial^2 x}{\partial r \partial f}$. By using $\frac{\partial x}{\partial f} = \frac{- \frac{\partial FOC}{\partial f}}{SOC}$, $ \frac{\partial FOC}{\partial f} = -\frac{r_1\mathrm{e}^{\frac{r_1x}{q}-{\Delta}r_1}}{q\left(\mathrm{e}^\frac{r_1x}{q}-1\right)^2}$ and $SOC = \frac{c r_1 e^{\frac{r_1x}{q}-\Delta r_1+\Delta r}}{q(e^\frac{r_1x}{q}-1)}$, I get

\begin{equation*}
    \frac{\partial x}{\partial f} =  \dfrac{\mathrm{e}^{-{\Delta}r}}{c\left(\mathrm{e}^\frac{r_1x}{q}-1\right)}
\end{equation*}

Which gives,
\begin{equation*}
   \frac{\partial^2 x}{\partial r \partial f} = -\dfrac{\Delta \mathrm{e}^{-{\Delta}r}}{c\left(\mathrm{e}^\frac{r_1x}{q}-1\right)}\leq 0
\end{equation*}

Thus,
\begin{equation*}
    \dfrac{\partial^2 D}{\partial r \partial f} \leq 0
\end{equation*}

For $r_1$,
\begin{equation*}
   \frac{\partial^2 x}{\partial r_1 \partial f} = -\dfrac{x \mathrm{e}^{\frac{r_1x}{q}-\Delta r}}{qc\left(\mathrm{e}^\frac{r_1x}{q}-1\right)^2} \leq 0
\end{equation*}

Thus,
\begin{equation*}
    \dfrac{\partial^2 D}{\partial r_1 \partial f} \leq 0
\end{equation*}

For $\Delta$,
\begin{equation*}
   \frac{\partial^2 x}{\partial \Delta \partial f} = -\dfrac{r \mathrm{e}^{-{\Delta}r}}{c\left(\mathrm{e}^\frac{r_1x}{q}-1\right)}\leq 0
\end{equation*}

Thus,
\begin{equation*}
    \dfrac{\partial^2 D}{\partial \Delta \partial f} \leq 0
\end{equation*}

\vspace{1cm}
\textbf{Proof of Proposition 4:}

\textbf{a)} 

\textbf{(i)}
By plugging in optimized order size in the objective function to get the total procurement cost
\begin{equation*}
    C(\Delta, f, ..) = \dfrac{\left(c\mathrm{e}^{{\Delta}r}x(.)+f\right)\mathrm{e}^{{-\Delta}r_1}}{1-\mathrm{e}^{-\frac{r_1x(.)}{q}}}
\end{equation*}

By using envelop theorem, I can get how the change in $r$ affects procurement costs\footnote{There is a closed form solution of the FOC. I can write equation \ref{eq:3} as
\begin{equation*}
    \mathrm{e}^{ax}-ax-b=0
\end{equation*}

Where, $a\equiv \frac{r_1}{q}$ and $b\equiv (\frac{fr_1}{cq\mathrm{e}^{\Delta r}}+1)$. The closed form solution of above equation is $x=-\dfrac{W(-\frac{1}{\mathrm{e}^b})+b}{a}$. Where, $W$ is a Lambert W function, which can solve equations with $x$ both in the base and the exponent. The Lambert W function shows that for positive $x$ values and the real number $a$ and $b$, there exists a unique $x^*$.}
\begin{equation*}
    \dfrac{\partial C(\Delta, f, ..)}{\partial r} = \dfrac{cx(.)\Delta \mathrm{e}^{\Delta r{-\Delta}r_1}}{1-\mathrm{e}^{-\frac{r_1x(.)}{q}}}
\end{equation*}

For $x>0$,
\begin{equation*}
    \dfrac{\partial C(\Delta, f, ..)}{\partial r} > 0
\end{equation*}

\textbf{(ii)}
Using the envelope theorem, I can understand how the change in $\Delta$ affects procurement costs.

\begin{equation*}
    \dfrac{\partial C(\Delta, f, ..)}{\partial \Delta} = \dfrac{c x(.)(r-r_1)\mathrm{e}^{{\Delta}(r-r_1)}-f r_1 \mathrm{e}^{-{\Delta}r_1} }{1-\mathrm{e}^{-\frac{r_1x(.)}{q}}}
\end{equation*}

The sign of the above derivative is ambiguous. For a clear effect, I need to consider the more realistic scenario behind the model. According to the model setup, for a firm of a developing country exporting to a developed country, $r>r_1$. I also know that production and financing costs are much higher than the per-shipment fixed costs, $cx\mathrm{e}^{\Delta r}>>f$. Thus, I get
\begin{equation*}
    \dfrac{\partial C(\Delta, f, ..)}{\partial \Delta} \geq 0
\end{equation*}

\textbf{(iii)}
By using envelop theorem, I can get how the change in $f$ affects procurement costs
\begin{equation*}
    \dfrac{\partial C(\Delta, f, ..)}{\partial f} = \dfrac{\mathrm{e}^{{-\Delta}r_1}}{1-\mathrm{e}^{-\frac{r_1x(.)}{q}}}
\end{equation*}

For $x>0$,
\begin{equation*}
    \dfrac{\partial C(\Delta, f, ..)}{\partial f} > 0
\end{equation*}

\textbf{(iv)}
By using envelop theorem, I can get how the change in $r_1$ affects procurement costs
\begin{equation*}
    \dfrac{\partial C(\Delta, f, ..)}{\partial r_1} =- \dfrac{\left(c\mathrm{e}^{{\Delta}r}x(.)+f\right)\mathrm{e}^{{-\Delta}r_1}\bigg[\Delta (1-\mathrm{e}^{-\frac{r_1x(.)}{q}})+\frac{x(.)}{q}\mathrm{e}^{-\frac{r_1x(.)}{q}}\bigg]}{(1-\mathrm{e}^{-\frac{r_1x(.)}{q}})^2}
\end{equation*}

For $x>0$,
\begin{equation*}
    \dfrac{\partial C(\Delta, f, ..)}{\partial r_1} < 0
\end{equation*}

\vspace{1cm}
\textbf{b)} 

\textbf{(i)} 
By taking derivative of $\dfrac{\partial C( f, ..)}{\partial \Delta}$ with respect to $f$
\begin{equation*}
    \dfrac{\big(c(r-r_1)\mathrm{e}^{\Delta (r-r_1)} \frac{\partial x}{\partial f}\big)(1-\mathrm{e}^{-\frac{r_1x(.)}{q}}-\frac{xr_1}{q}\mathrm{e}^{-\frac{r_1x(.)}{q}})+r_1\mathrm{e}^{-\Delta r_1}\big(\frac{fr_1}{q}\frac{\partial x}{\partial f}\mathrm{e}^{-\frac{r_1x(.)}{q}}-1+\mathrm{e}^{-\frac{r_1x(.)}{q}}\big)}{(1-\mathrm{e}^{-\frac{r_1x(.)}{q}})^2}
\end{equation*}

Defining $y\equiv \frac{xr_1}{q}$ and rearranging,
\begin{equation} \label{eqpc} 
    \dfrac{\partial^2 C}{\partial \Delta \partial f} = \dfrac{1}{(1-\mathrm{e}^{-\frac{r_1x(.)}{q}})^2}\bigg(\big(c(r-r_1)\mathrm{e}^{\Delta (r-r_1)} \frac{\partial x}{\partial f}\big)\big[1-\mathrm{e}^{-y}-y\mathrm{e}^{-y}\big]+r_1\mathrm{e}^{-\Delta r_1}\big[\mathrm{e}^{-y}(\frac{fr_1}{q}\frac{\partial x}{\partial f}+1)-1\big]\bigg)
\end{equation}

By using equation \ref{eqpc1}, $\frac{\partial x}{\partial f}\geq 0$. For $y=0$, the first square bracketed terms are zero and the second square bracket term is positive. It turns out that for $0< y < 1$, the two square bracketed terms are always positive. Thus, for $0< \frac{xr_1}{q} < 1$ and $r>r_1$, the equations \ref{eqpc1} and \ref{eqpc} give 
\begin{equation*}
    \dfrac{\partial^2 C}{\partial \Delta \partial f} \geq 0
\end{equation*}

Now for $r$,
\begin{equation*}
    \dfrac{\partial^2 C}{\partial r \partial f} = -\dfrac{\frac{\partial x(.)}{\partial r}\mathrm{e}^{-\Delta r_1-\frac{x(.)r_1}{q}}}{(1-\mathrm{e}^{-\frac{r_1x(.)}{q}})^2}
\end{equation*}

As $\frac{\partial x(.)}{\partial r}\leq 0$,
\begin{equation*}
    \dfrac{\partial^2 C}{\partial r \partial f} \geq 0
\end{equation*}

\textbf{(ii)} 
Now for $r_1$,
\begin{equation*}
    \dfrac{\partial^2 C}{\partial r_1 \partial f} = -\dfrac{[r_1\frac{\partial x(.)}{\partial r_1}+x(.)]\frac{1}{q}\mathrm{e}^{-\Delta r_1-\frac{x(.)r_1}{q}}+\Delta \mathrm{e}^{-\Delta r_1}(1-\mathrm{e}^{-\frac{r_1x(.)}{q}})}{(1-\mathrm{e}^{-\frac{r_1x(.)}{q}})^2}
\end{equation*}

For $x>0$ and a small value of $r_1$, the change of shipping size for the importer lending rate cannot be larger than the shipping size itself. Thus, square bracketed term is positive as $\frac{\partial x(.)}{\partial r}\leq 0$. I get,
\begin{equation*}
    \dfrac{\partial^2 C}{\partial r_1 \partial f} < 0
\end{equation*}

\subsection*{External Financing for per-Shipment Costs} \label{a2}

In this alternative setup, I consider an extreme assumption regarding per-shipment cost. Assume the per-shipment cost is financed immediately after getting the order .\footnote {This is unlikely if the firms do not borrow money upfront for per-shipment costs; they pay much later. If the lead time (delivery time) is long, the firm will not immediately borrow the per-shipment cost after receiving the order.} This scenario is useful if the shipping time is much larger than the order processing and production time \citep{hasan2017supply}. This setup can be imagined as the bank funding working capital and per-shipment cost. After realizing the importer's revenue, the bank deducts loaned funds, including financing and shipping costs, and only handovers the profit to the exporter.  

The objective function is
\begin{equation*}
 \min_{x} \dfrac{\mathrm{e}^{{\Delta}r}\left(cx+f\right)}{1-\mathrm{e}^{-\frac{r_1x}{q}}}
\end{equation*}
 
Taking derivative with respect ot $x$,
\begin{equation*} 
   \dfrac{\left(cq\mathrm{e}^\frac{r_1x}{q}-cr_1x-fr_1-cq\right)\mathrm{e}^{\frac{r_1x}{q}+{\Delta}r}}{q\left(\mathrm{e}^\frac{r_1x}{q}-1\right)^2}
\end{equation*}

The optimal order size satisfies (by setting the first-order condition equal to zero), 
\begin{equation} \label{eqship} 
    cq\mathrm{e}^\frac{r_1x^*}{q}=cr_1x^*+fr_1+cq
\end{equation}

By using equation \ref{eqship}, for $x>0$ and $r_1>0$, the cost is minimized as the second-order condition is
\begin{equation*}
    \dfrac{r_1\left(\left(cr_1x+fr_1-2cq\right)\mathrm{e}^\frac{r_1x}{q}+cr_1x+fr_1+2cq\right)\mathrm{e}^{\frac{r_1x}{q}+{\Delta}r}}{q^2\left(\mathrm{e}^\frac{r_1x}{q}-1\right)^3}> 0
\end{equation*}

From equation \ref{eqship}, I find $x$ has a positive relationship with $f$. The per-shipment fixed cost's effect on shipping frequency does not deviate from the original setup. However, in this alternative setup, financing cost at the origin, $r$, and delivery time, $ \Delta $, do not affect order size or shipping frequency. For importing country financing cost $r_1$, I find a similar effect to the original model. Thus, it will be an empirical question of whether this alternative or original model works for a country's trade scenario.

\subsection{Appendix-B} \label{b}

\subsection*{Per-Shipment Size and Export Volume} \label{b1}
The theoretical model also predicts the effect of financing costs on the order size. To estimate this effect, I use the following model:
\begin{equation}\label{ch2sols} 
    \begin{multlined}
     ln\ v_{igmjt}=\alpha +spline1+spline2+ \beta_1 ln\ f_{jt}+ \beta_2 ln\ distance_{j} + \beta_3 spline1 \times ln\ distance_{j}\\ + \beta_4 spline2 \times ln\ distance_{j}+ \beta_5 r^{exp}_{t}+ \beta_6 r^{imp}_{jt} + \theta X_{jt} + \lambda_{i}+ \eta_{g}+ \phi_{m} +\xi_{t}+ \epsilon_{igmjt}
    \end{multlined}
\end{equation}

Where, $ln\ v_{igjmt}$ is either the log per-shipment value or the log export volume (by value or weight) of Bangladeshi firm $i$ for good $g$ with transport mode $m$ to country $j$ in year $t$. All other notations are defined in the previous models. The per-shipment value captures the order size. As more frequent shipments mean small-sized orders, I expect the opposite sign of the previous models for all the beta coefficients. I estimate the log-linear model \ref{ch2sols} by using OLS. For trade volume, negative $\beta_1$ means higher per-shipment costs lead to less export volume. However, for per-shipment value, positive $\beta_1$ means higher per-shipment costs lead to a larger shipping size. Similarly, positive $\beta_5$ and $\beta_6$ mean greater financing costs lead to larger-sized shipments.

From the theoretical understanding, the financing cost's effect on the per-shipment value is the opposite of the shipping frequency. To find the empirical evidence for this effect, I run regressions as per the estimation equation \ref{ch2sols}. Table \ref{ch2size} presents the results. Column 1 uses the per-shipment value as the dependent variable. As expected, the sign flipped for the lending rates. But per-shipment fixed costs still show a negative coefficient contradicting the predictions. Column 2 captures the gravity approach and shows a significant negative effect of per-shipment costs on the export volume \citep{kropf2014fixed,hornok2015administrative}. Finally, in column 3, I use export weight instead of export volume and find similar coefficients in sign, significance, and magnitude compared to column 2.

\pagebreak
\begin{table}[H] 
  \centering
  \tiny
  \singlespacing
  \renewcommand\thetable{12}
  \caption{Shipping Size}
    \begin{tabular}{lccc}
    \toprule
          & (1)   & (2)   & (3) \\
     & Ln Per-shipment Value    & Ln Export Volume    & Ln Export Volume \\
      &     & (Value)    & (Weight) \\
    \midrule
        
    Ln per-shipment cost & -0.157*** & -0.292*** & -0.304*** \\
          & [0.016] & [0.025] & [0.026] \\
    Ln distance & -1.973*** & -2.651*** & -2.775*** \\
          & [0.112] & [0.137] & [0.140] \\
    Spline1 & -13.648*** & -18.368*** & -20.058*** \\
          & [0.914] & [1.118] & [1.157] \\
    Spline2 & -30.982*** & -43.419*** & -45.169*** \\
          & [1.434] & [1.691] & [1.714] \\
    
    Spline1$\times$ Ln distance & 1.217*** & 1.659*** & 1.843*** \\
          & [0.095] & [0.118] & [0.122] \\
    
    Spline2$\times$ Ln distance & 3.361*** & 4.722*** & 4.910*** \\
          & [0.154] & [0.181] & [0.184] \\
    Importer interest rate    & 0.039*** & 0.003 & 0.004 \\
          & [0.008] & [0.012] & [0.013] \\
    
    Importer interest rate$\times$Exporter interest rate & -0.002*** & -0.000 & -0.000 \\
          & [0.001] & [0.001] & [0.001] \\
    Ln GDP  & 0.178*** & 0.343*** & 0.347*** \\
          & [0.004] & [0.006] & [0.006] \\
    Ln GDP per capita & 0.118*** & 0.237*** & 0.217*** \\
          & [0.010] & [0.017] & [0.017] \\
    Island & -0.141*** & -0.231*** & -0.234*** \\
          & [0.022] & [0.033] & [0.034] \\
    Landlocked & -0.103*** & -0.258*** & -0.257*** \\
          & [0.022] & [0.033] & [0.034] \\
    common religion & 0.814*** & 1.170*** & 1.334*** \\
          & [0.080] & [0.126] & [0.127] \\
    Common legal origin & -0.268*** & -0.257*** & -0.186*** \\
          & [0.019] & [0.031] & [0.032] \\
    Colony & 0.253*** & 0.454*** & 0.433*** \\
          & [0.029] & [0.048] & [0.050] \\
    Constant & 23.182*** & 25.902*** & 24.686*** \\
          & [1.047] & [1.296] & [1.329] \\
   \midrule
    Observations & 371,871 & 371,871 & 371,869 \\
    R-squared & 0.592 & 0.530 & 0.589 \\
   
    \bottomrule

    \bottomrule
    \end{tabular}%
  \label{ch2size}%
\end{table}%
\vspace{-1em}
\footnotesize{Note: OLS regressions of the estimation equation \ref{ch2sols}. The dependent variable is the log of per-shipment value, and the log of export volume (by value and weight). Exporter by product by mode by year fixed effects is used in all regressions. Standard errors are robust and clustered with exporter clusters. ***$p < 0.01$; **$p < 0.05$; *$p < 0.1$.}

\end{document}